\documentclass{elsarticle}
\usepackage{hyperref,color}
\usepackage{todonotes}
\usepackage{ulem}
\usepackage{amssymb}
\usepackage{booktabs}
\usepackage{chemformula}
\setchemformula{radical-radius=1pt}
\RequirePackage{xspace}
\DeclareFontFamily{OT1}{pzc}{}
\DeclareFontShape{OT1}{pzc}{m}{it}{<-> s * [1.200] pzcmi7t}{}
\DeclareMathAlphabet{\mathpzc}{OT1}{pzc}{m}{it}
\usepackage{amsfonts}

\journal{Nuclear Instruments and Methods A}

\usepackage{multirow}
\newcommand{\doserate}{\mathcal{R}}
\newcommand {\SCSNeightone} {SCSN$-$81~}
\newcommand{\unit}[1]{\ensuremath{\textrm{\,#1}}\xspace}
\newcommand {\pDi}{\mathpzc{D^{-1}}}

\begin{document}

\begin{frontmatter}

\title{Dose rate effects in radiation-induced changes to phenyl-based polymeric scintillators}

\author[umd]{C.~Papageorgakis\corref{cor}}
\ead{cpapag@umd.edu}
\author[umd2]{M.~Al-Sheikhly}
\author[umd]{A.~Belloni}
\author[umd]{T.K.~Edberg}
\author[umd]{S.C.~Eno}
\author[fnal]{Yongbin~Feng}
\author[umd]{Geng-Yuan~Jeng\fnref{fn1}}
\author[penn]{Abraham~Kahn}
\author[umd]{Yihui~Lai}
\author[umd]{T.~McDonnell\fnref{fn2}}
\author[umd]{Ameer~Mohammed\fnref{fn3}}
\author[umd]{C.~Palmer}
\author[umd]{Ruhi~Perez-Gokhale\fnref{fn2}}
\author[umd]{F.~Ricci-Tam\fnref{fn4}}
\author[umd]{Zishuo~Yang}
\author[umd]{Yao~Yao\fnref{fn5}}

\cortext[cor]{Corresponding author}
\fntext[fn1]{currently at ArcPoint Forensics, Tampa FL}
\fntext[fn2]{currently at Rohde \& Schwarz, Columbia MD}
\fntext[fn3]{currently at Virginia Polytechnic Institute and State University}
\fntext[fn4]{currently at Sigmoid Health, Santa Clara CA}
\fntext[fn5]{currently at U.C. Davis}

\affiliation[umd]{ 
    organization={Dept. Physics, U. Maryland},
    city={College Park},
    state={MD},
    country={USA}
}
\affiliation[umd2]{ 
    organization={Dept. Materials Science and Engineering, U. Maryland},
    city={College Park},
    state={MD},
    country={USA}
}
\affiliation[penn]{ 
    organization={Dept. Physics \& Astronomy, U. Pennsylvania},
    city={Philadelphia},
    state={PA},
    country={USA}
}
\affiliation[fnal]{ 
    organization={Fermi National Accelerator Laboratory},
    city={Batavia},
    state={IL},
    country={USA}
}

\begin{abstract}
Results on the effects of ionizing radiation on the signal produced by plastic scintillating rods manufactured by Eljen Technology company are presented for various matrix materials, dopant concentrations, fluors (EJ-200 and EJ-260), anti-oxidant concentrations, scintillator thickness, doses, and dose rates.  The light output before and after irradiation is measured using an alpha source and a photomultiplier tube, and the light transmission by a spectrophotometer.  Assuming an exponential decrease in the light output with dose, the change in light output is quantified using the exponential dose constant $D$.  The $D$ values are similar for primary and secondary doping concentrations of 1 and 2 times, and for antioxidant concentrations of 0, 1, and 2 times, the default manufacturer's concentration. The $D$ value depends approximately linearly on the logarithm of the dose rate for dose rates between 2.2\unit{Gy/hr} and 100\unit{Gy/hr} for all materials. For EJ-200 polyvinyltoluene-based (PVT) scintillator, the dose constant is approximately linear in the logarithm of the dose rate up to 3900 Gy/hr, while for polystyrene-based (PS) scintillator or for both materials with EJ-260 fluors, it remains constant or decreases (depending on doping concentration) above about 100 Gy/hr. The results from rods of varying thickness and from the different fluors suggest damage to the initial light output is a larger effect than color center formation for scintillator thickness $\leq1$~cm. For the blue scintillator (EJ-200), the transmission measurements indicate damage to the fluors.  We also find that while PVT is more resistant to radiation damage than PS at dose rates higher than about 100\unit{Gy/hr} for EJ-200 fluors, they show similar damage at lower dose rates and for EJ-260 fluors.
\end{abstract}


\end{frontmatter}

\section{Introduction}
When particle detectors based on plastic scintillator are exposed to ionizing radiation, their light output decreases with absorbed dose.  For reasonably low doses, the dependence of the signal strength on the dose $d$ can be approximated by an exponential
\begin{equation}
L(d)  = L_0 \exp(-d/D),
\label{eqn:exp}
\end{equation}
where $L(d)$ is the signal after irradiation, $L_0$ is the signal before irradiation, and $D$ is the ``dose constant", a numeric parameter whose value depends on the scintillator geometry, the specific scintillator used, environmental factors, and on the dose rate $\frac{d}{dt}(d) \equiv \doserate$.  Larger values of $D$ correspond to greater radiation tolerance.

Because their exposures are at low $\doserate$ (typically a few $\times 10^{-3}-10^{0}$\unit{Gy/hr}), the value of $D$ for plastic scintillators in calorimeters at current and future collider experiments at CERN's Large Hadron Collider is challenging to measure, as low $\doserate$ exposures are of necessity of long duration and therefore can be expensive.  The CMS collaboration has recently~\cite{Sirunyan_2020} reported measurements at $\doserate$s between a few $\times 10^{-3}$ and a few $\times 10^{4}$\unit{Gy/h} for plastic scintillator tiles (either \SCSNeightone from the Kuraray Corporation\footnote{Kuraray, Ote Center Building, 1-1-3, Otemachi, Chiyoda-ku, Tokyo 100-8115, Japan} or BC-408 from the Bicron division of the Saint-Gobain Corporation\footnote{Saint Gobain Corp, Les Miroirs, 18, Avenue d'Alsace, 92400 Courbevoie, France}), with an embedded Y$-$11 wavelength-shifting fiber (Kuraray).  The wavelength-shifting fiber is connected to a clear plastic fiber that leads to a photodetector.  Their results show a power law dependence of $D$ on $\doserate$ for $\doserate$ less than a few $10$'s of Gy/hr.  At higher $\doserate$s, due to systematic uncertainties, the behavior is consistent either with a continued power law dependence or a constant $D$.  Other recent measurements of radiation damage to plastic scintillator for particle detectors are reported in  Refs.~\cite{Jivan_2015,Pedro_2019}.  Two recent reviews of radiation damage in plastic scintillator are Refs.~\cite{sauli,dubna}.  For a comprehensive review of plastic scintillators, see Ref.~\cite{hamel}

Several studies of the $\doserate$ dependence of radiation damage in plastic scintillators have been published~\cite{Biagtan1996125,34504,Wick1991472,289295,173180,173178,Giokaris1993315,1748-0221-11-10-T10004,gillen}. However, previous measurements of $\doserate$ effects in scintillators without wavelength-shifting fibers were limited to high $\doserate$s and were unable to differentiate different potential causes. In this paper, we present results for $D$ for plastic scintillating rods manufactured by the Eljen Technology company\footnote{Eljen Technology, 1300 W. Broadway, Sweetwater, Texas 79556, United States} using polystyrene (PS) or polyvinyltoluene (PVT) as the matrix, several dopant concentrations, two different fluors (EJ-200 and EJ-260), various scintillator thicknesses, and three anti-oxidant concentrations for $\doserate$s from 2.2\unit{Gy/hr} to 3900\unit{Gy/hr}. These variations allow exploration of potential causes. In addition, our results separate the contributions to the CMS results from the plastic scintillator and the other components in the tiles and their readout.

\section{Mechanisms and effects of radiation-induced changes to phenyl-based polymeric scintillators}

Common plastic scintillator consists of a matrix (often PS or PVT) containing primary and secondary fluors. Ionizing radiation excites the matrix.  This excitation can be transferred to the primary fluor via the F{\"o}rster mechanism~\cite{forster}, which dominates at primary concentrations above approximately 1\%~\cite{birks}, or radiatively in the UV, which dominates at low concentrations. The excitation is transferred radiatively from the primary to the secondary, which de-excites via the emission of visible light.
 
Radiation affects the light output of scintillators mainly through damage to the polymer substrate.  The effects depend on the chemical structure, the degree of crystallinity of the irradiated polymers, and the total dose and dose rate. Dissolved oxygen in the amorphous part can also play an important role in the radiation chemistry of the polymers~\cite{cloughquant,clough2,GILLEN1995149} both during and after irradiation. The atactic polymers used in our work are amorphous, so there is no effect due to the degree of crystallinity.  For amorphous PS and PVT, the phenyl group on their backbone chains strongly decreases radiation effects~\cite{RadChem}, as its $\pi$ structure provides excellent protection to the polymer chains. The photoelectric and Compton electrons resulting from gamma radiolysis of PS and PVT react very rapidly with the $\pi$ structure of these phenyl groups, producing anion radicals. It is expected that the phenyl anions convert rapidly to C-centered radicals through a protonation reaction to produce benzyl-type radicals~\cite{RadChem2}. In amorphous materials, such as atactic PS and PVT, these benzyl-type and toluenyl C-centered radicals undergo the various competing reactions depending on the presence of oxygen and the $\doserate$. 

Decrease in light output can come from two sources: either through the formation of so-called ``color centers" that absorb light emitted from the secondary fluor in transit from its creation point to the photodetector, or through a decrease in the initial production of light. Decrease of the initial light production can occur through fluor destruction or alteration, and through absorption of light in transmission between the primary and the secondary fluor. Additionally, radiation damage can enhance the de-excitation of the matrix via mechanisms that do not lead to light production (``quenching"). For example, oxygen can quench the initial light, as the excitation can transfer to it instead of to the primary fluor~\cite{sauli,HORSTMANN1993395}. 
In addition, since the Birks' constant, important for highly ionizing particles, depends on the density of the primary dopant and on interactions between excited substrate molecules~\cite{POSCHL2021164865}, its value could change.

Radicals can absorb visible light, and generally absorb more strongly at shorter wavelengths~\cite{sauli}.  Radicals are the source of the so-called temporary damage, which decreases with time after the radiation exposure as the radicals recombine with each other to produce stable molecules. After irradiation, \ch{R-C^.} radicals in the amorphous region continue to react with oxygen and undergo cross-linking reactions.  However, the \ch{R-C^.} radicals have a much longer lifetime.  Free radicals centered on the backbone of the polymer chain in the crystalline region transfer to the amorphous region via the hydrogen-hopping mechanism.  This process is called annealing.  After annealing, the remaining color centers are referred to as permanent damage. Radicals can also undergo recombination reactions (including cross-linking reactions) during irradiation. The non-standard bonds to the matrix material can form permanent bonds that can absorb in the visible, acting as color centers.

The rate of radical production during irradiation goes as~\cite{Busjan199989}
\begin{equation}
\frac{d[Y]}{dt}=gQ\doserate-k[Y]^2,
\label{eqn:exp2}
\end{equation}
where $[Y]$ is the density of radicals, $g$ is the radiation-chemical yield, $Q$ is the scintillator density, and k is the reaction constant for the decay of the radical.  The first term represents the creation of radicals, while the second term represents two unterminated radicals recombining to neutralize (``second-order" termination). At short times, when the second term is small compared to the first, integration yields a radical density that is proportional to dose: $Y=gQd$. If the radical is not neutralized by e.g. oxidation, then the second term grows with time, and eventually, a steady state is reached, when the two terms are equal. In this case, the radical density becomes constant with time, is no longer proportional to dose, and deviations from the expected exponential behavior described by equation~\ref{eqn:exp} occur. When the first term dominates, Eq.~\ref{eqn:exp} becomes:
\begin{equation}
L(d) = L_0 \exp(-gQd\sigma l),
\label{eqn:exp4}
\end{equation}
where $\sigma$ is the cross section for absorption of light by the color centers, and $l$ is the light's path length through the scintillator to the photodetector.  This would indicate $D^{-1}=gQ \sigma l$, and so $D$ would scale with $l^{-1}.$

\subsection{The reactions of the C-centered radicals in the absence of oxygen}

Radiolytically produced C-centered radicals of PS and PVT undergo crosslinking reactions in the absence of oxygen. However, the steric effect of the phenyl group on the backbone of the PS and PVT chains impede crosslinking reactions, leading to lower crosslinking radiation yield.  For example, absorption of 2000\,eV gamma energy produces only one crosslink~\cite{RadChem}. The polystyrene free radicals can also react with H-atoms, which are produced during the radiolysis, leading to their disappearance on a longer timescale.

\subsection{The reactions of the C-centered radicals in the presence of oxygen}

While high $\doserate$ enhances the crosslinking reactions, low $\doserate$ and the presence of oxygen promote the oxidation processes.

The depth $z_0$ for oxygen diffusion into a rectangular slab of plastic is~\cite{cunliffe}
\begin{equation}
z_0^2=\frac{2 \, M \, C_0}{\Upsilon \, \doserate}=\frac{2 \, M \, S \, P}{\Upsilon \, \doserate},
\label{eqn:z0}
\end{equation}
where $M$ is the diffusion coefficient for oxygen, $C_0$ is the oxygen concentration at the matrix's surface on the matrix side, $\Upsilon~(=gQ)$ is the specific rate constant of active site formation, $S$ is the oxygen solubility, and $P$ is the external oxygen pressure. In general, these parameters depend on temperature~\cite{difftemp}. There is an abrupt transition between areas with and without oxygen.  The oxygen concentration in the oxidized regions is almost uniform~\cite{cloughPS}. For PS rods with a thickness of 1\unit{cm}, oxygen permeates the entire sample for $\doserate$s below (roughly, depending on the plastic preparation and environment) 1.6\unit{Gy/h}~\cite{cloughPS,Wick1991472}. For thicknesses of 0.4\unit{cm}, 0.6\unit{cm}, and 0.8\unit{cm}, the corresponding permeation $\doserate$s are below 10\unit{Gy/h}, 4.4\unit{Gy/h}, and 2.5\unit{Gy/h} respectively. For $\doserate$s above this value, polymer oxidation will occur only in the region permeated by oxygen, contributing to a $\doserate$ dependence of the damage to the scintillator.

The rate of polymer oxidation is~\cite{clough2, bolland1,clough1,cloughquant, GILLEN1995149}
\begin{equation}
K(C(x,t))= -\frac{C_{1} \, C(x,t)}{1+C_{2} \, C(x,t)} ,
\label{eqn:diff2}
\end{equation}
where $-K(C(x,t))$ is the rate at which oxygen diffuses through the polymer bulk, $x$ is the position relative to the surface of the material where the rate is being measured, and $C(x,t)$ is the position-dependent concentration of oxygen within the matrix. The constants $C_1$ and $C_2$ depend on the kinematics of the chemical reactions. The constant $C_1$ is proportional to the square root of the $\doserate$ for bimolecular reactions (leading to a dose-rate effect) and to $\doserate$ for unimolecular reactions (no dose-rate effect because integration yields a proportionality to dose).

The decay kinetics of the PS- benzyl-type and PVT toluenyl C-centered radicals can be summarized as follows:
\begin{equation}
    -\frac{d[\ch{R-C^.}]}{dt} = k_1[\ch{O2}][[\ch{R-C^.}]+2k_2[\ch{R-C^.}]^2]
\end{equation}
where $k_1$ and $k_2$ represent the reaction rate constants of the C-centered radicals with oxygen and the bi-molecular second-order cross-linking reactions respectively. At a sufficiently high $\doserate$, the rate of free radical production $\frac{d[\ch{R-C^.}]}{dt}$ is high enough to lead to an enhancement of the cross-linking reaction. Therefore, at very high $\doserate$, cross-linking competes well with the diffusion rate of oxygen into the bulk of the irradiated atactic PS and PVT. Also note that, at room temperature, atactic PS and PVT are regarded as oxygen barriers, and oxygen diffusion is very low. So at high $\doserate$, the crosslinking reactions are dominant in the bulk.  However, at the surfaces of the irradiated samples, PS and PVT samples where the \ch{O2} is available, their radiolytically produced carbon-centered radical \ch{R-C^.} react very rapidly with oxygen to give rise to the formation of corresponding peroxyl radicals~\cite{cloughquant,clough2,GILLEN1995149,polym12122877}.

The peroxyl radicals undergo bimolecular reactions to produce unstable tetraoxide intermediates 
\begin{equation}
    \ch{R-CO2^. + R-CO2^. -> ROOOOR}
\end{equation}
The \ch{ROOOOR} intermediates undergo various decomposition reactions, producing relatively stable oxides such as organic and hydroperoxides (ROH and ROOH). These oxides can absorb strongly in the ultraviolet, leading to reduced transfer of light to the secondary fluor.  Their absorption cross section decreases with increasing wavelength. Note that \ch{R-CO_2^.} can abstract an H atom from the backbone of the neighboring molecules, producing more \ch{R-C^.}, and initiating a short-chain reaction.
\begin{equation}
    \ch{R-C^. + R-CH -> R-C^. + R-CO2H}
\end{equation}
where R-CH is either PS or PVT
In this study, we used glassy atactic (100\% amorphous) PS and PVT.  Hence, the permeability of \ch{O2} through PS and PVT is very small~\cite{PhysPolymer}.  So, the traces of oxygen in the bulk are consumed by irradiation at a much faster rate than replenished by its permeability from outside. Hence it is expected that the crosslinking reactions are the predominant in the bulk of the PS and PVT irradiated samples, and the oxidation reactions take place mainly on the surfaces of these samples at high dose rates. Our results show the presence of oxidation products in the bulk too. This is because our samples are relatively thin, and hence the presence of oxygen in the bulk. In addition, our very low dose rates impede the crosslinking reactions and enhances the reaction of polystyrene radicals with oxygen.

\section{Sample and irradiation details}
\begin{figure}[hbtp]
\centering
\includegraphics[width=0.8\textwidth]{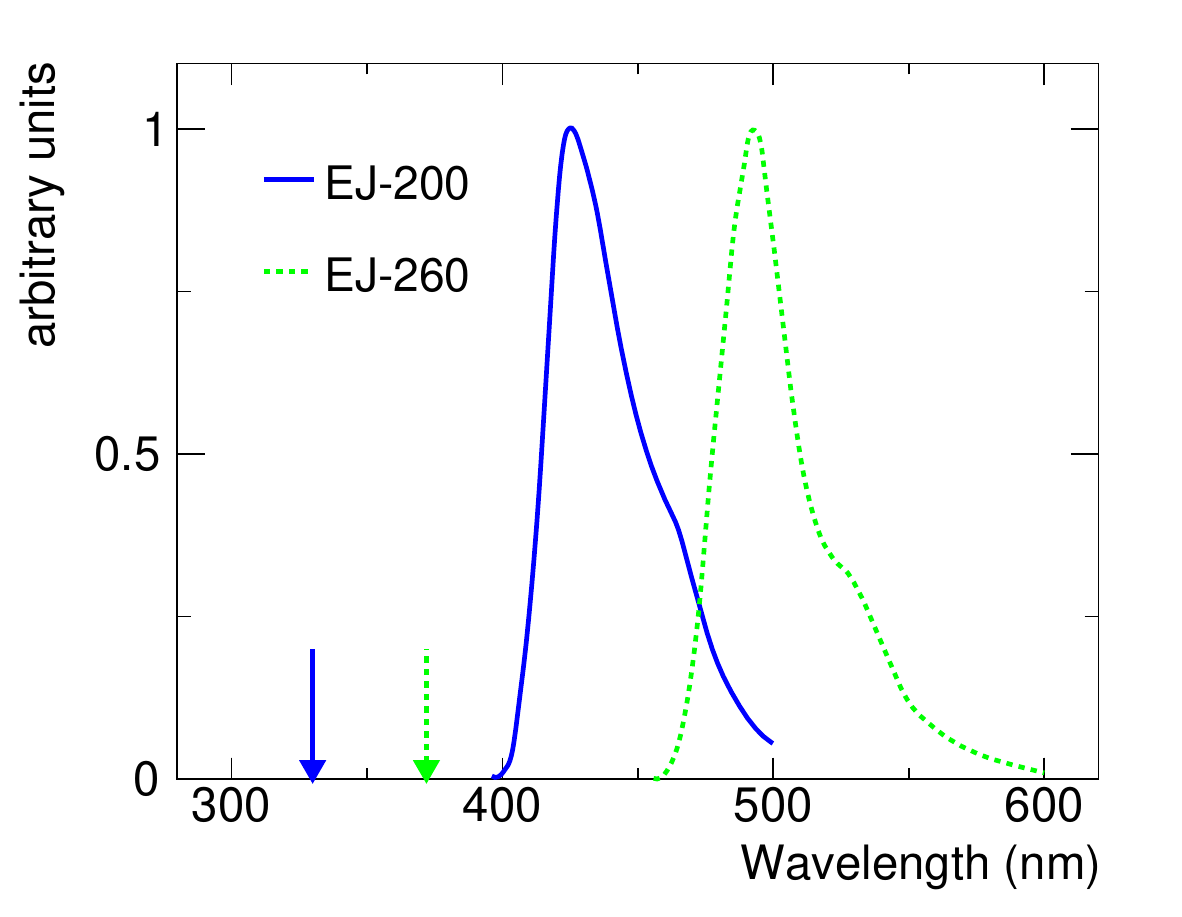}
\caption{Emission spectra from the Eljen Technology website for EJ-200 (blue solid) and EJ-260 (green dot).  The arrows represent the emission maxima for the primary for EJ-200 (blue solid) and EJ-260 (green dot).
}
\label{fig:emission}
\end{figure}

Our scintillator samples are in the form of rectangular rods, 5~cm long and 1~cm wide, with thicknesses of 0.4\unit{cm}, 0.6\unit{cm}, 0.8\unit{cm}, and 1.0\unit{cm}.  All faces are diamond milled. The rods, supplied by Eljen Technology, contain either the primary and secondary fluors used in EJ-200 (a blue scintillator with p-Terphenyl as the primary and \ch{POPOP}-type as the secondary) or EJ-260 (a green scintillator whose fluors are proprietary to Eljen Technology), with either PS or PVT as the plastic matrix.  The secondary fluor emission spectra from the Eljen website for EJ-200 and EJ-260 are shown in Fig.~\ref{fig:emission}. For the EJ-200 fluors, the wavelength of maximum emission for the primary fluor in PS or PVT is around 320--350\unit{nm}, and the region most important for secondary emission is 410--480\unit{nm}. The primary fluor emission maximum for EJ-260 is 372\unit{nm}~\cite{maxwell}, and the secondary fluor emits primarily between 475 and 550\unit{nm}. Some of the rods had double the normal concentration of the primary dopant, and some had double concentration for the secondary dopant.  Some had no anti-oxidant included, and some had twice the nominal anti-oxidant concentration. Fig.~\ref{fig:hd} [top] shows a photograph of some of the rods.  Rods were acquired in several purchases or gifts over a period of several years. Lists of used samples are given in the Tables that can be found in \ref{app:A}.

Irradiations were performed at four facilities with gamma ray sources. High $\doserate$ irradiations, with $\doserate$s between 80.6 and 3900\unit{Gy/hr}, were performed at the National Institute of Standards and Technology, Gaithersburg, MD and at Sandia National Laboratories using \ch{^{60}Co} sources. Low $\doserate$ irradiations, at 3.0, 3.1, and 9.8\unit{Gy/hr}, were performed at Goddard Space Flight Center using a \ch{^{60}Co} source, and at 2.2\unit{Gy/hr} at the GIF++ facility~\cite{gif} at CERN using a \ch{^{137}Cs} source. The irradiations were done at room temperature.  The humidity and oxygen pressure were not measured or controlled.

The irradiations at Sandia National Laboratories were subject to long multi-hour pauses. The longest pauses lasted 1-3 days while the full annealing time for this material is approximately 1 month. Scintillator chemistry is expected to be affected by annealing due to radical recombination and deeper oxygen penetration when irradiation is paused. For this reason, data from these irradiations are not used for dose constant calculations and they are included only in a comparison between light output and dose (Fig.~\ref{fig:expdose}). 

The uncertainties on the accumulated dose and $\doserate$ were $\pm 10\%$ for irradiations performed at Goddard Space Flight Center and GIF++, and $\pm1.3\%$ (95\% confidence level) for irradiations performed at the National Institute of Standards and Technology. Accumulated doses (in water equivalent) ranged from 12.6\unit{kGy} to 70\unit{kGy}.  Irradiations were performed at various times over a period of several years. The doses were chosen to approximately halve the light yield.

\section{Measurement technique}
\begin{figure}[hbtp]
\centering
\includegraphics[width=0.6\textwidth]{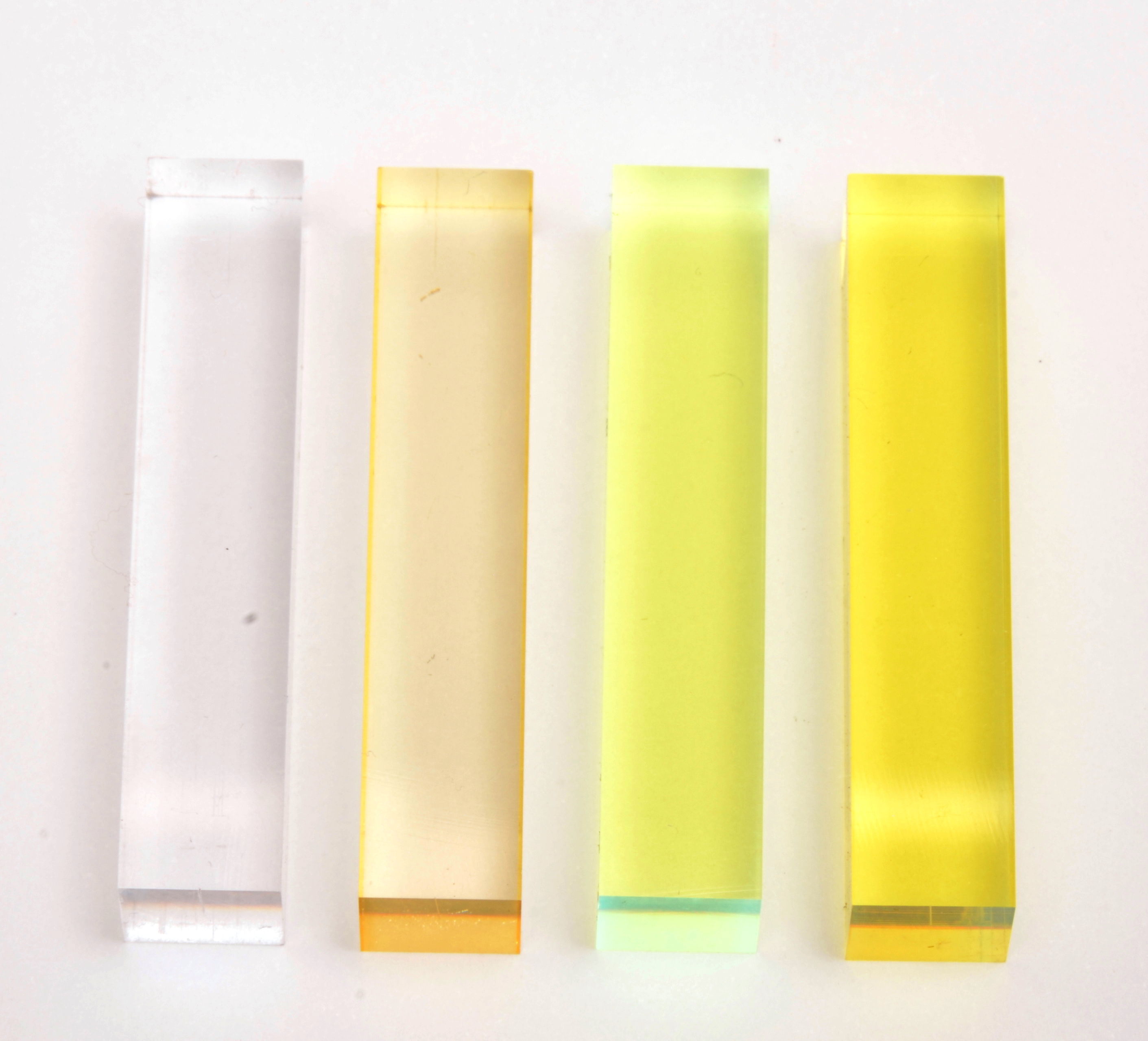}
\includegraphics[width=0.8\textwidth]{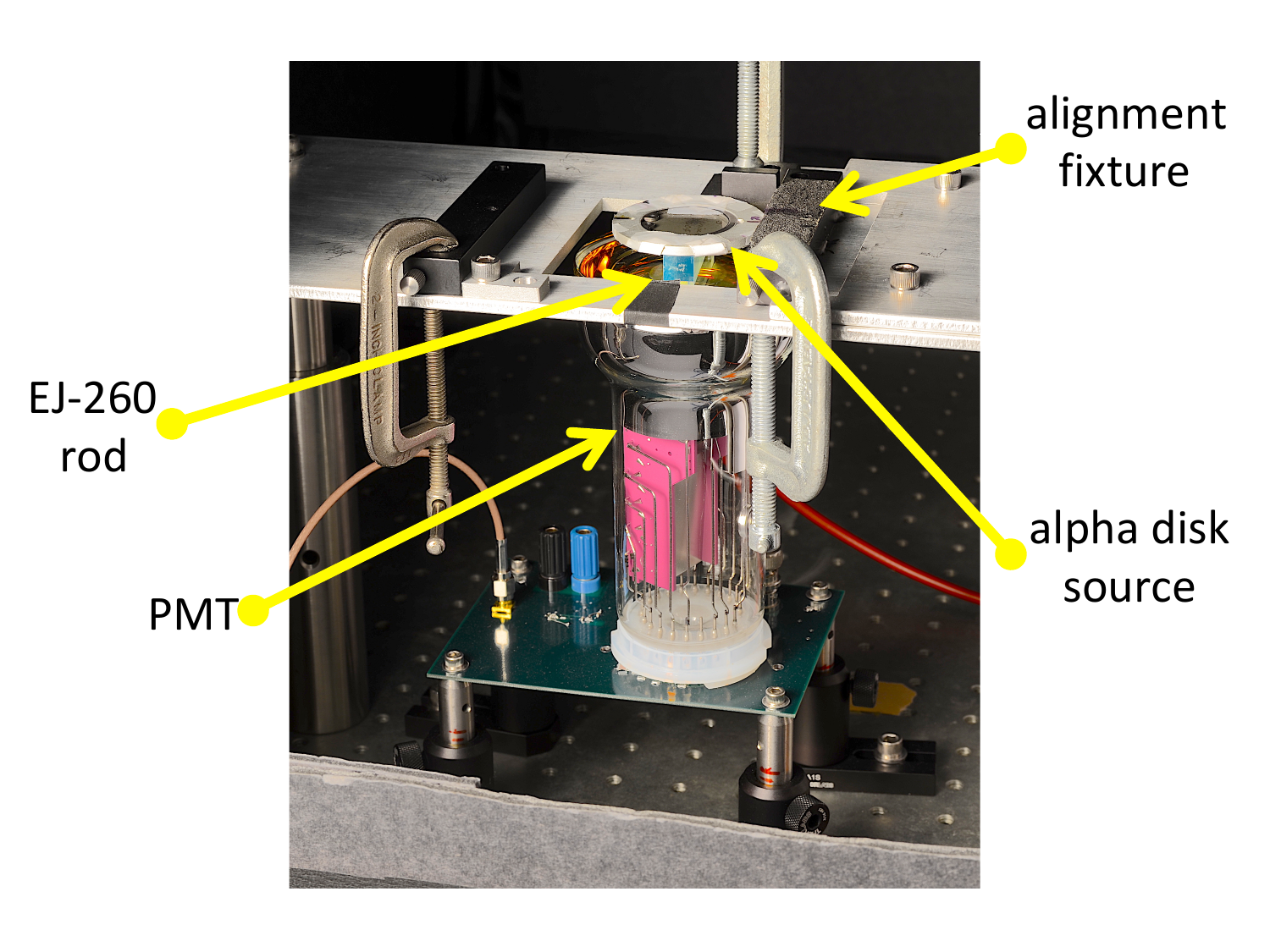}
\caption{[top] A photograph of selected rods, all with nominal doping.  From left to right: unirradiated EJ-200, EJ-200 irradiated to 500~kGy at 11~kGy/hr, unirradiated EJ-260, EJ-260 irradiated to 500~kGy at 11~kGy/hr.
[bottom] Apparatus for measurements with an alpha source.  In use, the apparatus is enclosed in a light-tight box.
}
\label{fig:hd}
\end{figure}

\begin{figure}[hbtp]
\centering
\includegraphics[width=0.75\textwidth]{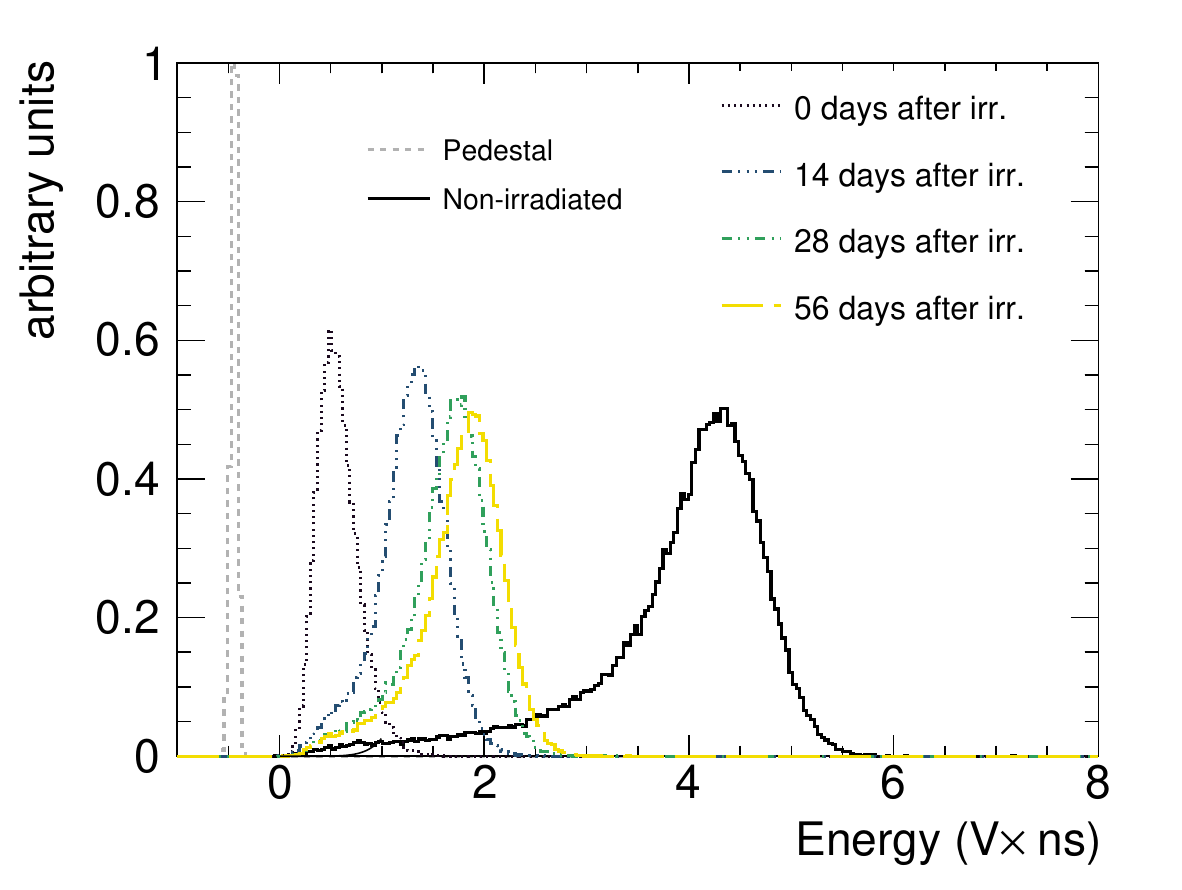}
\includegraphics[width=0.75\textwidth]{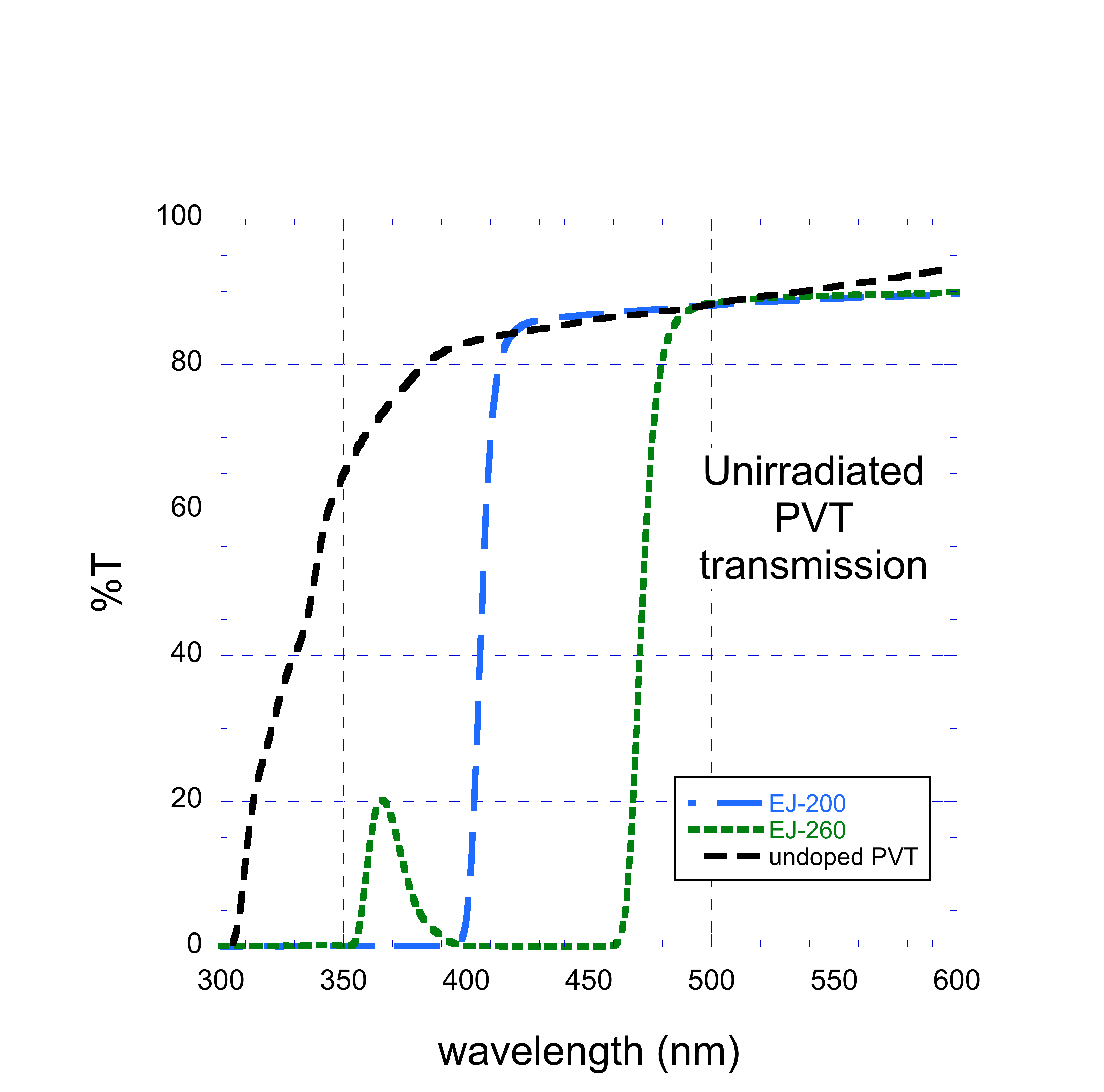}
\caption{[top] Measured energy spectrum from a rod irradiated at 3640\unit{Gy/hr} to a dose of 70\unit{kGy} when excited using an alpha source at various times since irradiation, showing the annealing process.
[bottom] Typical transmission spectra for rods with nominal fluor and anti-oxidant concentrations and with a PVT matrix, with no fluors (black, medium dashes), EJ-260 fluors (green, short dashes), and for EJ-200 fluors (blue, long dashes).
}
\label{fig:typical}
\end{figure}

The light output from the rods is measured before and after irradiation using an alpha source (\ch{^{239}Pu}, 80~nCi) and a Hamamatsu R6091 photomultiplier tube, as shown in Fig.~\ref{fig:hd} [bottom]. The penetration depth of its 5.156 MeV alpha in PS is 0.037\unit{mm}~\cite{astar}. Each rod was placed on the photomultiplier tube, operated at +1,700\unit{V}, and the source was placed on the rod.  An alignment fixture ensured reproducibility of the alignment of the three pieces.  The measurements were made using a Tektronix oscilloscope model TDS7104, with a charge integration window of 100~ns. The measurements used in subsequent plots, except where explicitly noted, occurred after annealing was complete.  Our measurements are therefore of the permanent damage. Figure~\ref{fig:typical} [top] shows alpha spectra at various times after irradiation for a rod, illustrating the annealing process.

After pedestal subtraction, the distribution is fit to a Gaussian around the peak, and its mean is used as a measure of the light yield. Before and after a series of rod measurements, we measure a standard reference rod to calibrate any photomultiplier gain drift and to validate the performance of the equipment.  Each rod is measured multiple times. The systematic uncertainty in the light output is dominated by imperfect registration of the photomultiplier, rod, and alpha source.  This systematic uncertainty was estimated from the variation in repeated light yield measurements of a given rod and is $\pm 0.5$\%. Other sources of systematic uncertainty in the precision of the signal measurement for an individual rod are negligible.  Systematic uncertainties due to effects of environmental factors and manufacturing tolerance on variation between rods are on the order of a few percent.

Transmission measurements as a function of wavelength were taken using a Varian Cary 300 spectrophotometer. Fig.~\ref{fig:typical} [bottom] shows some results of typical transmission measurements. The position of the edge in transmission for the scintillating rods corresponds to the end of the absorption spectrum for the secondary fluor.
 
In order to facilitate comparison of the spectra from samples with different doses, a pseudo-inverse of $D$, $\pDi$, is calculated as a function of wavelength:
\begin{equation}
\pDi = \frac{\ln(T_o)-\ln(T_f)}{d}
\label{eqn:exp3}\end{equation}
where $T_o$ and $T_f$ are the transmission as a function of wavelength before and after irradiation, respectively.

\section{Results}

\begin{figure}[hbtp]
\centering
\includegraphics[width=0.49\textwidth]{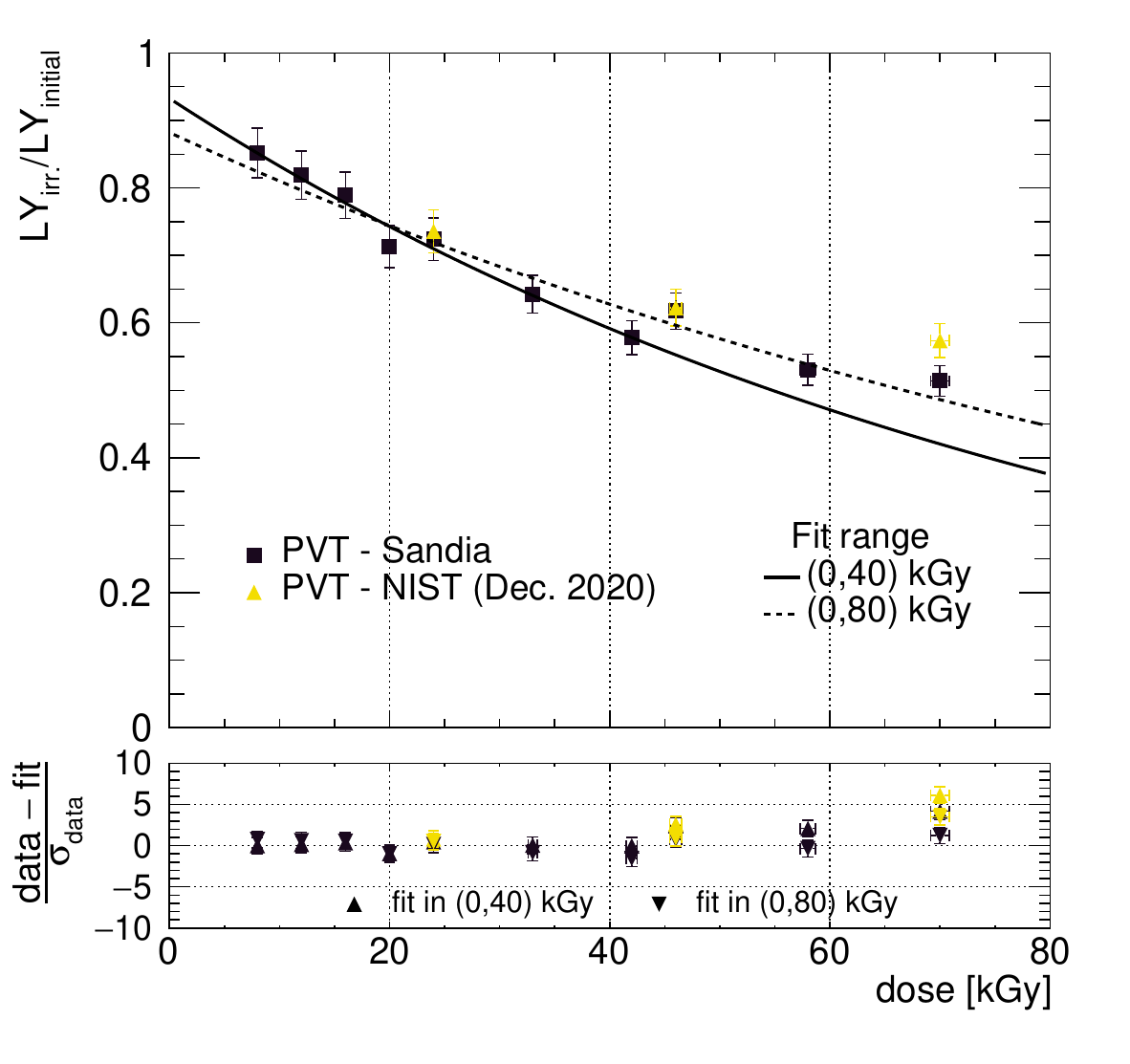}
\includegraphics[width=0.49\textwidth]{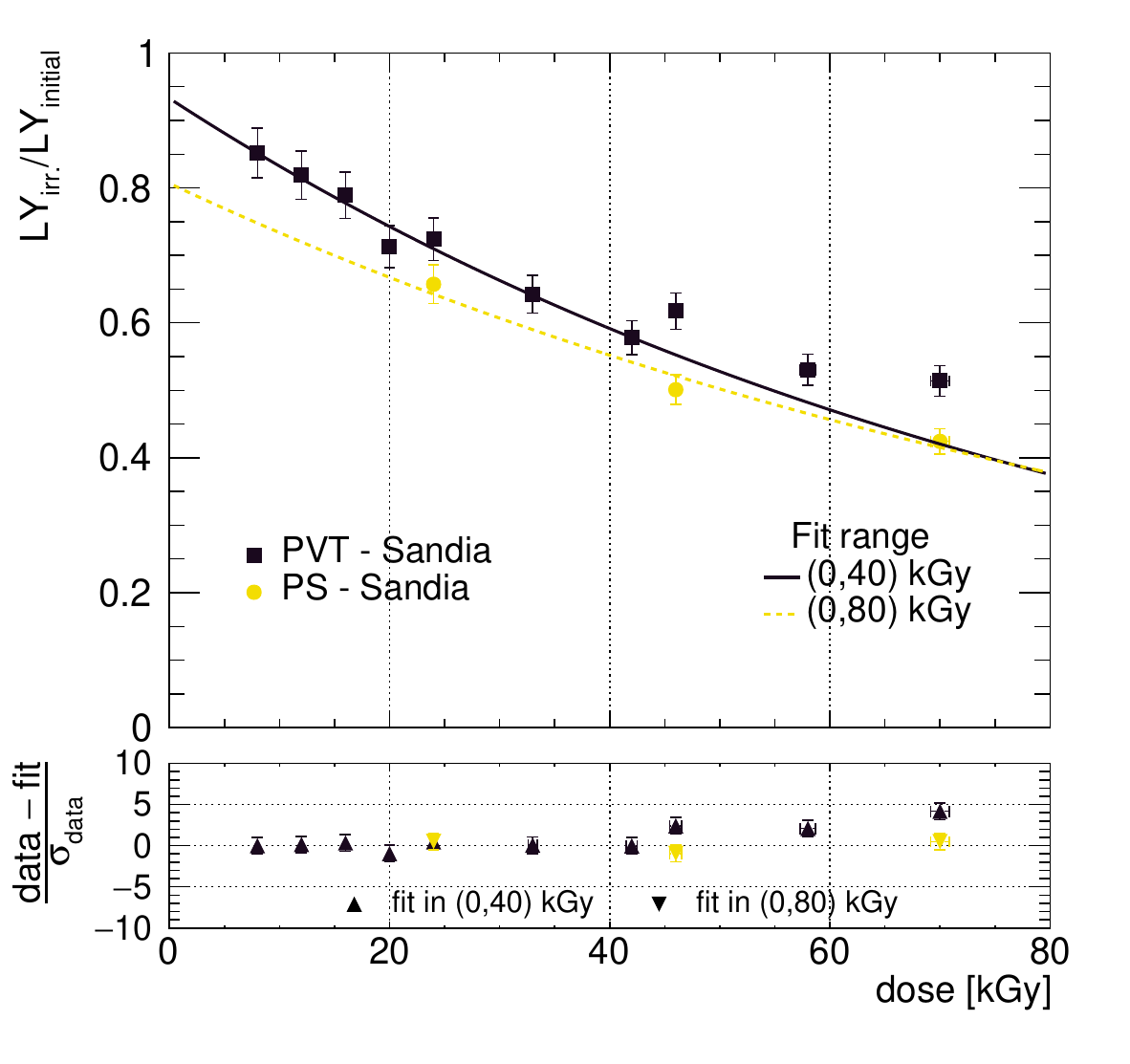}
\caption{[left] Ratio of the light output of rods after receiving a dose $d$ to their unirradiated light output, versus $d$. The rods' matrix is PVT and the irradiation dose rate was 0.46\unit{kGy/hr}.  The different colors correspond to different irradiation dates and facilities. Rods irradiated at Sandia National Laboratories and NIST are represented with black and yellow markers, respectively.  The lines are fits of the Sandia data to exponentials, one using only the lower dose data and the other using all data.  Uncertainties are dominated by systematic uncertainties.
[right] Same, comparing PS and PVT. All rods included in this plot have been irradiated at Sandia National Laboratories.
}
\label{fig:expdose}
\end{figure}

Figure~\ref{fig:expdose} shows as a function of $d$ the ratio of the light output of a rod before and after irradiation.  The rod's matrix is PVT and the irradiation $\doserate$ was 460\unit{Gy/hr}. Each point corresponds to a separate rod. Some of the rods presented in this figure experienced pauses of 1-3 days with the full annealing time for this material being approximately 1 month.  The results at this $\doserate$ are well described by an exponential for doses below 40\unit{kGy}.  Above this, some saturation may be occurring.  Since many of our results are based on irradiation of 70\unit{kGy} (see \ref{app:A}), this may indicate the values are an underestimate of the damage for lower doses.  For the CMS detector in HL-LHC running, doses up to a few~\unit{kGy} are expected in the scintillator part of the high granularity calorimeter~\cite{hgcaltdr}.
 
\begin{figure}[hbtp]
\centering
\includegraphics[width=0.9\textwidth]{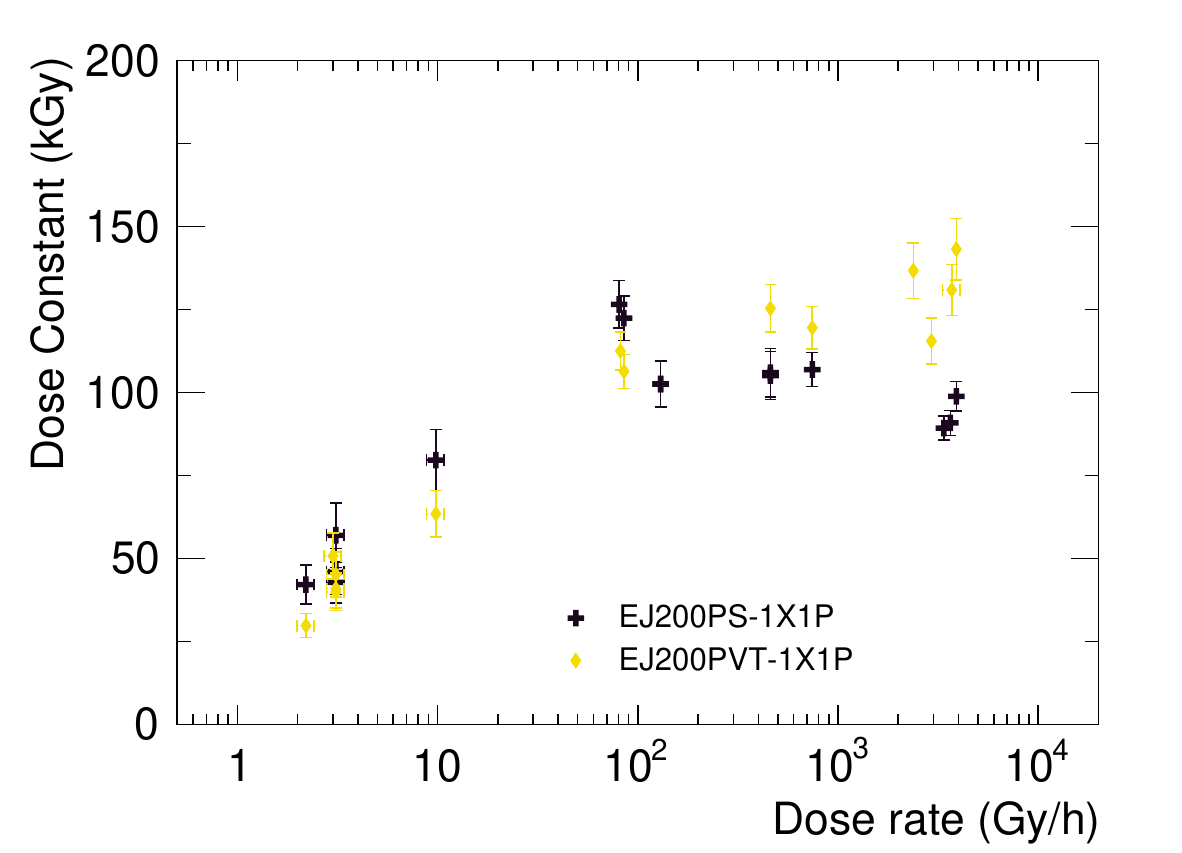}
\caption{The value of $D$ versus $\doserate$ for scintillator rods with the EJ-200 fluors at the manufacturer's nominal concentration and with nominal antioxidant concentration, for PVT (blue markers) and PS (black markers).
}
\label{fig:2001x1palpha}
\end{figure}

Figure~\ref{fig:2001x1palpha} shows $D$ versus $\doserate$ for scintillator rods with the EJ-200 fluors at the manufacturer's nominal concentration and with nominal antioxidant concentration, for both PS and PVT matrices. The values of $D$ for the two matrices are similar for $\doserate$s between 2 and 100\unit{Gy/hr}, and are approximately linear in the logarithm of the $\doserate$. For $\doserate>$100\unit{Gy/hr}, $D$ is larger at larger $\doserate$ for PVT while for PS it is smaller or constant at larger $\doserate$.  The scatter in the data reveal an uncontrolled systematic, perhaps related to changes in manufacturing over the period of the purchases or due to uncontrolled environmental variables. The radiation hardness of PS and PVT is similar at low $\doserate$.  At high $\doserate$, PVT is more radiation tolerant for EJ-200 (blue).

\begin{figure}[hbtp]
\centering
\includegraphics[width=0.9\textwidth]{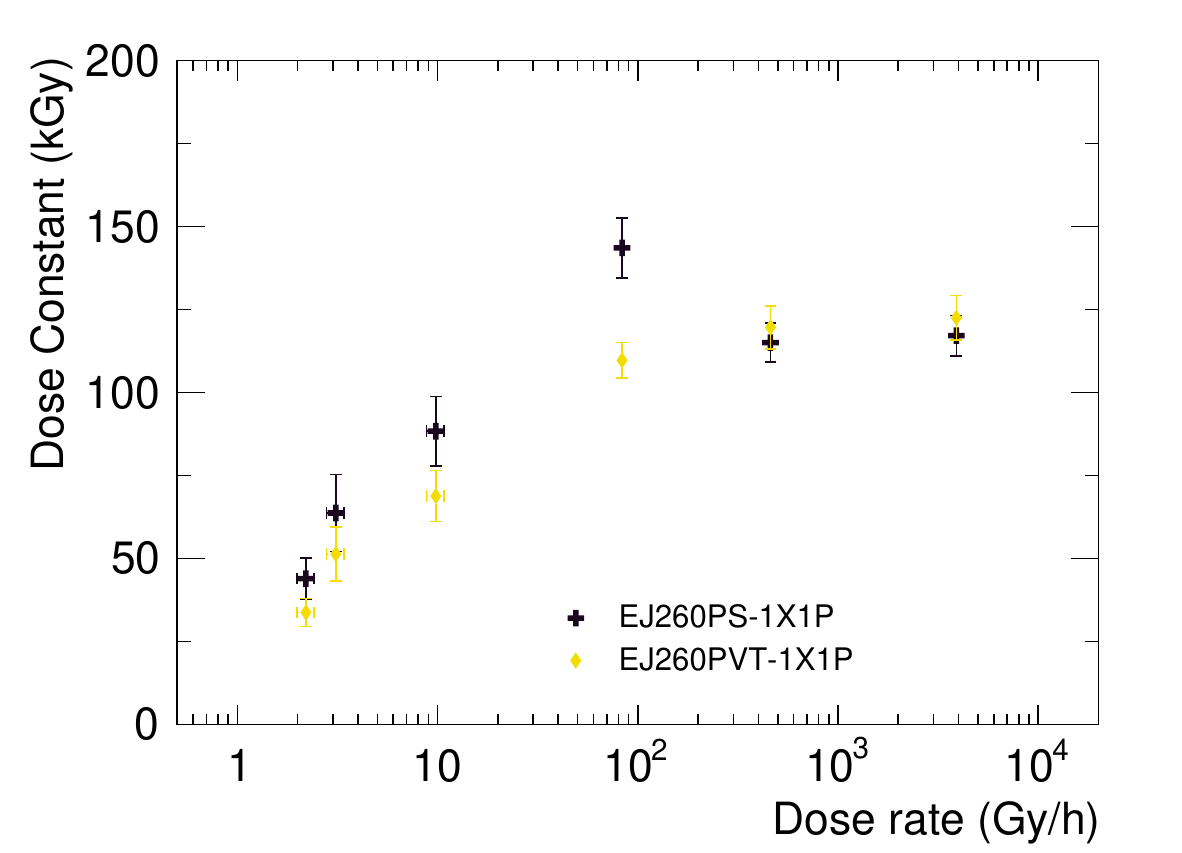}
\caption{The value of $D$ versus $\doserate$ for scintillator rods with the EJ-260 fluors at the manufacturer's nominal concentration and with nominal antioxidant concentration, for PS and PVT matrices. 
}
\label{fig:2601x1palpha}
\end{figure}

Figure~\ref{fig:2601x1palpha} shows $D$ versus $\doserate$ for scintillator rods with the EJ-260 fluors at the manufacturer's nominal concentration and nominal antioxidant concentration, for both PS and PVT matrices.  The results are similar to those for the EJ-200 fluors at $\doserate$s below about 100\unit{Gy/h}.  At higher $\doserate$s, the results for PVT are similar to those of EJ-200, while the radiation resistance of PS is improved.

\begin{figure}[hbtp]
\centering
\includegraphics[width=0.8\textwidth]{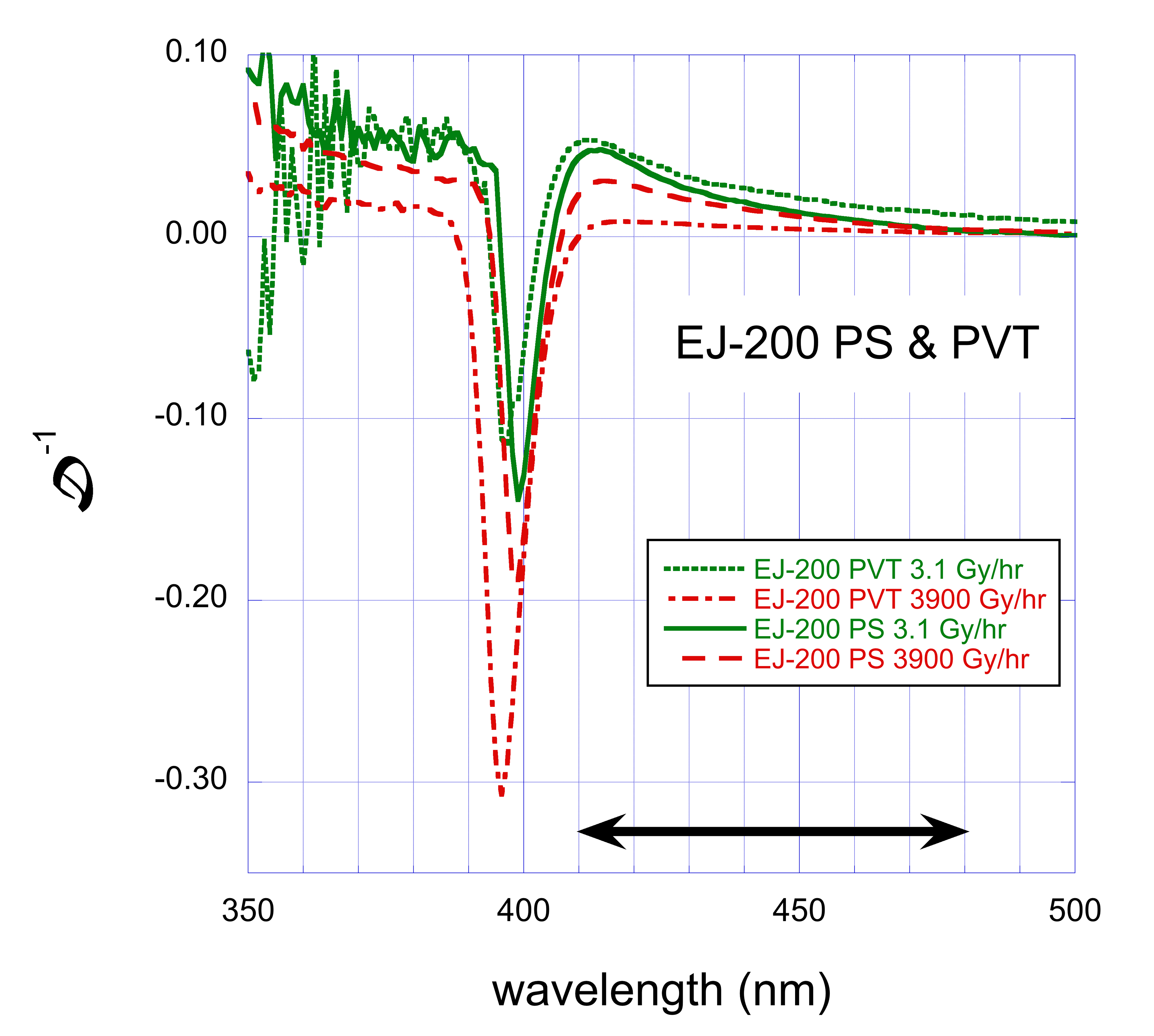}
\includegraphics[width=0.8\textwidth]{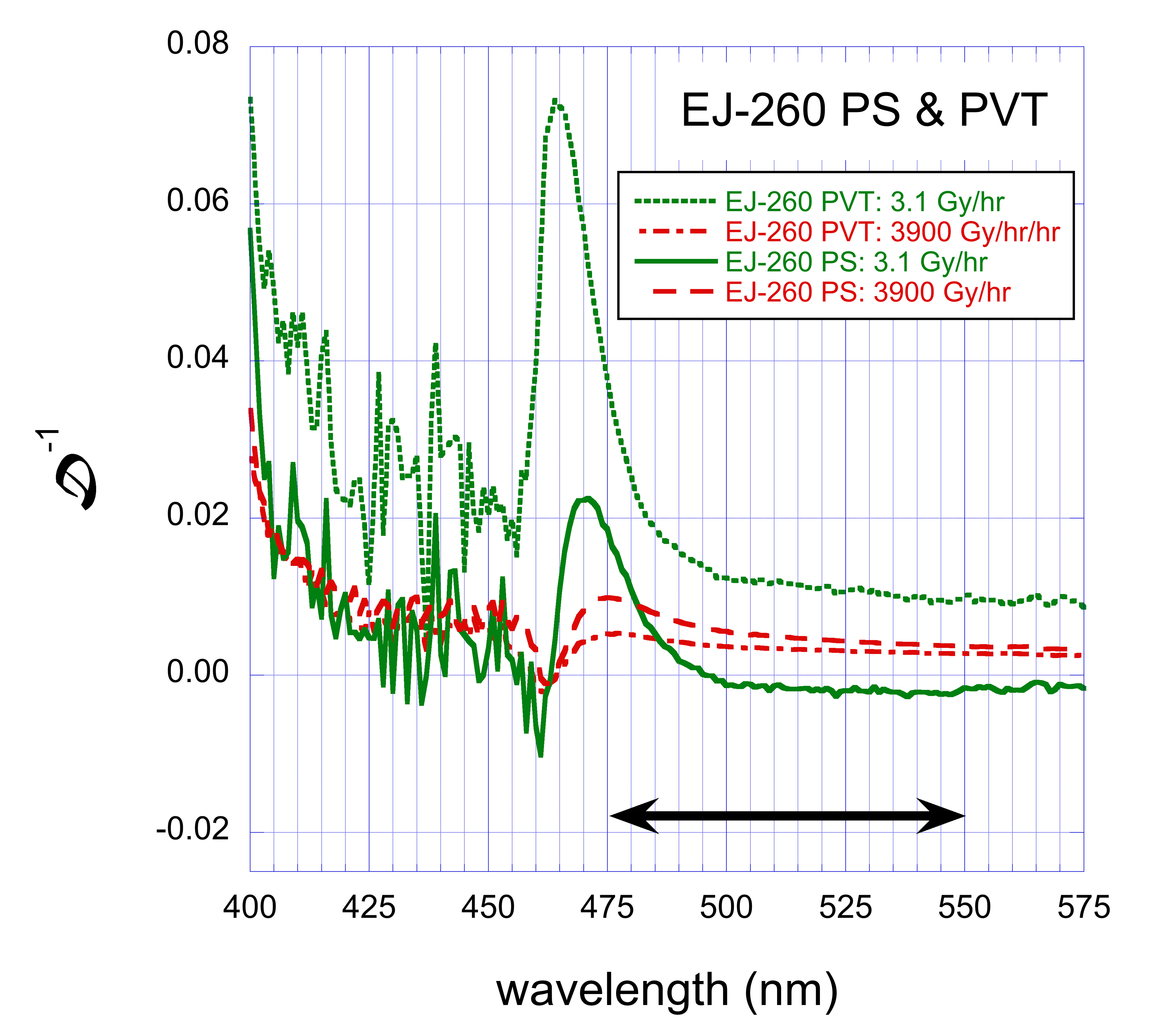}
\caption{[top] The value of $\pDi$ for scintillator with the EJ-200 fluors, for both PS and PVT, at $\doserate$s of 3.1\unit{Gy/hr} to a total dose of 12.6\unit{kGy} and 3900\unit{Gy/hr} to a total dose of 70\unit{kGy}.
[bottom] The value of $\pDi$ for scintillator with the EJ-260 fluors, for both PS and PVT, at $\doserate$s of 3.1\unit{Gy/hr} to a total dose of 12.6\unit{kGy} and 3900\unit{Gy/hr} to a total dose of 70\unit{kGy}.
In both plots, the emission range for the secondary fluor is indicated by a double arrow. 
}
\label{fig:1x1pabs}
\end{figure}

\begin{figure}[hbtp]
\centering
\includegraphics[width=0.76\textwidth]{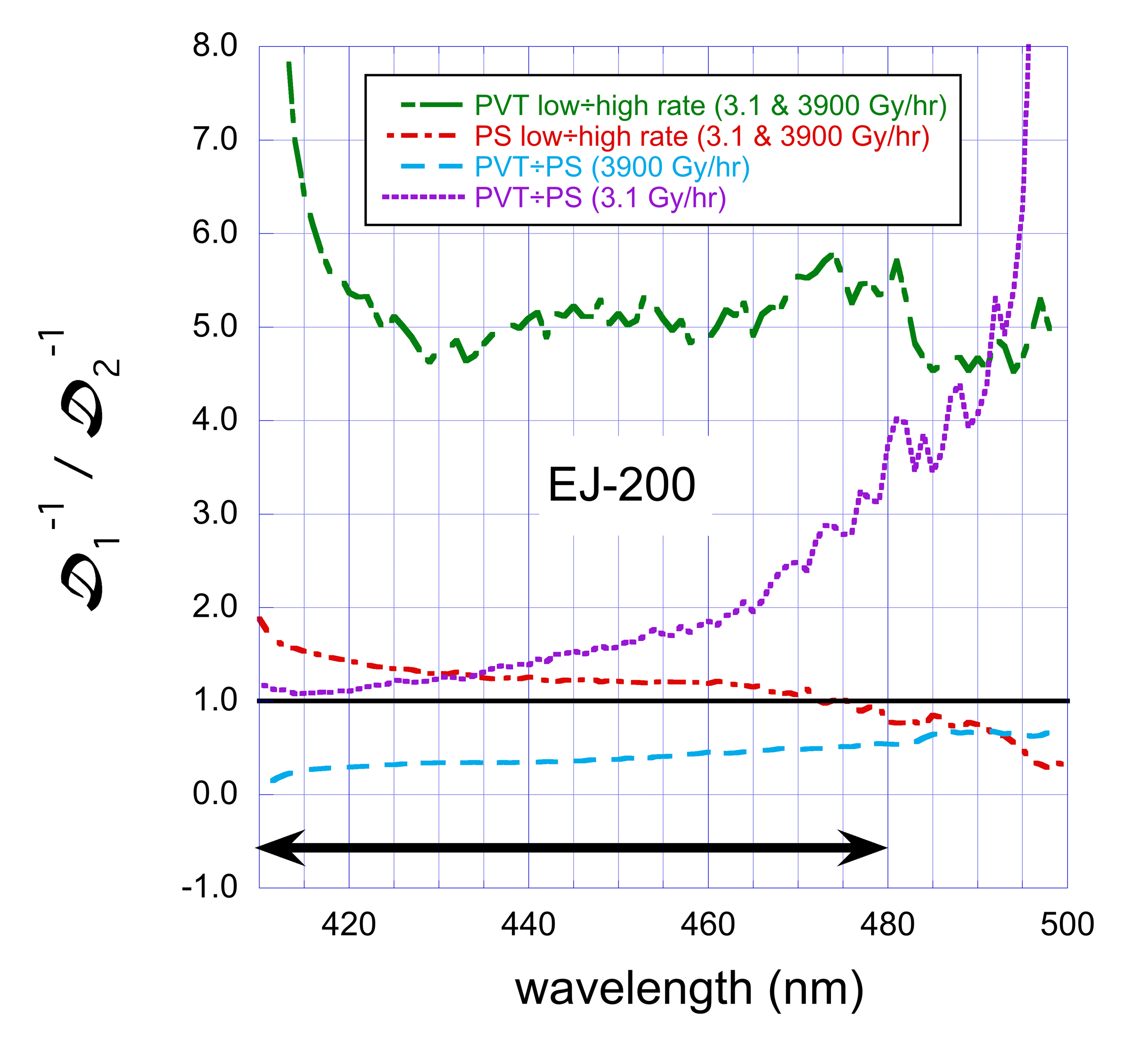}
\includegraphics[width=0.76\textwidth]{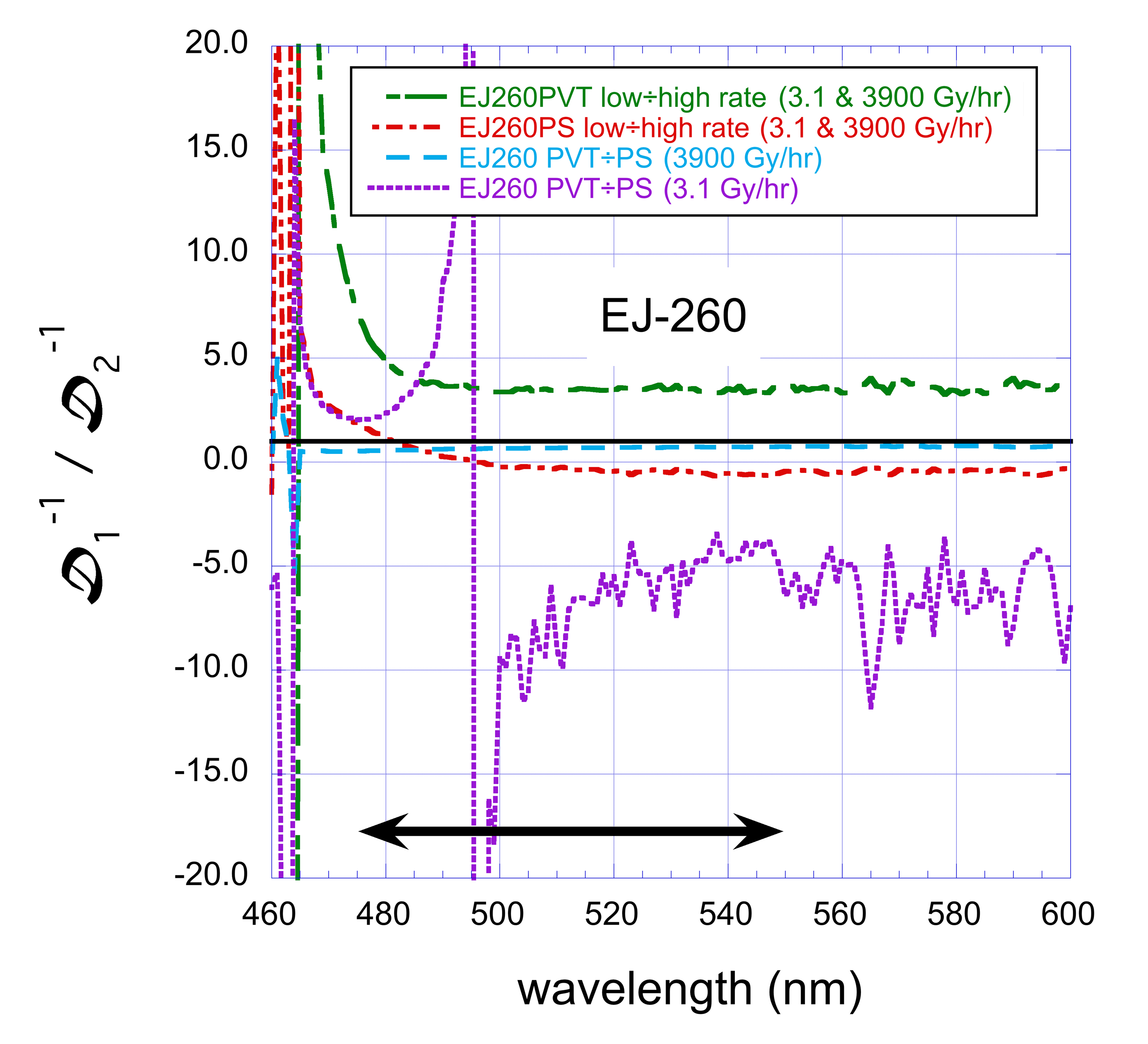}
\caption{Ratios $\mathpzc{D_{1}^{-1}}/\mathpzc{D_{2}^{-1}}$ are plotted versus wavelength, where $\mathpzc{D_{1}^{-1}}$ and $\mathpzc{D_{2}^{-1}}$ are values of $\pDi$ corresponding to different cases of irradiations or irradiated materials. The precise cases are defined separately for each ratio. Two different irradiations are included, a high $\doserate$ one at 3900\unit{Gy/hr} (total dose: 70\unit{kGy}) and a low $\doserate$ one at 3.1\unit{Gy/hr} (total dose: 12.6\unit{kGy}).
[top] EJ-200 for PVT at low $\doserate$ divided by high $\doserate$ (green dash), PS at low $\doserate$ divided by high $\doserate$ (red dot-dash), PVT divided by PS both at low $\doserate$ (purple dot), PVT divided by PS both at high $\doserate$ (red dot-dash)
[bottom] EJ-260 for PVT at low $\doserate$ over high $\doserate$ (green dash), PS at low $\doserate$ over high $\doserate$ (red dot-dash), PVT over PS both at low $\doserate$ (purple dot), PVT over PS both at high $\doserate$ (red dot-dash)
In both plots, the emission range for the secondary fluor is indicated by a double arrow. The solid black line indicates a value of 1.0.}
\label{fig:1x1pabsr}
\end{figure}

Figure~\ref{fig:1x1pabs} shows the $\pDi$ results for the PS and PVT versions of the EJ-200 and EJ-260 scintillators for $\doserate$s of 3.1\unit{Gy/hr} and 3900\unit{Gy/hr}. Larger positive values correspond to more damage. Negative values of $\pDi$ are consistent with destruction of the fluor, resulting in less light self-absorption. For EJ-200, the emission and absorption spectra for the secondary fluor overlap in the range 370\unit{nm}$<\lambda<$ 3900\unit{nm}. Figure~\ref{fig:1x1pabsr} shows ratios of $\pDi$ values for high and low $\doserate$, and for PS and PVT.  For EJ-200, in the region of secondary emission above 410\unit{nm}, there is more damage at low $\doserate$ than at high $\doserate$.  The damage is larger at lower wavelengths as expected from previous results~\cite{sauli}.  The values at low $\doserate$ for PS and PVT are similar. The value of $\pDi$ is negative in the region where the absorption and emission bands of the primary overlap, indicating some destruction of the secondary fluor.

For EJ-260, the damage was smaller, and the rods were too thin to allow accurate measurement in this case.  There is indication of color center formation just below 475\unit{nm}, which is at the start of the region of the primary emission.  In both PS and PVT, this feature did not appear in EJ-200.  These color centers formed more at low $\doserate$ than high, and more in PVT than in PS at low $\doserate$, and at about the same rate in PS and PVT at high $\doserate$.  Absorption from the color centers is low enough in wavelength, though, that it would not affect the light output very much. The ratios show that, except for the comparison between PVT and PS for EJ-200, the effects are roughly independent of wavelength.

\section{Thickness, fluor concentration, and anti-oxidant concentrations}
Changes in light output due to color center formation can be distinguished from decrease in initial light production through the dependence of $D$ on the rod thickness. 
If the color center formation is independent of whether or not the region is oxidized, D will scale as $l^{-1}$.  If the color centers form predominantly in the oxidized regions, this scaling will only occur once the entire sample is permeated with oxygen,
and will be and independent of thickness above this value, as the effective thickness is $2 z_0$.
For PS, at our lowest $\doserate$, oxygen permeates the rod for thicknesses below 0.6\unit{cm}. If the damage is to initial light production, $D$ would not depend on the thickness. Because all our rods are thin, the color center density must be large to produce measurable light attenuation.

\begin{figure}[hbtp]
\centering
\includegraphics[width=0.9\textwidth]{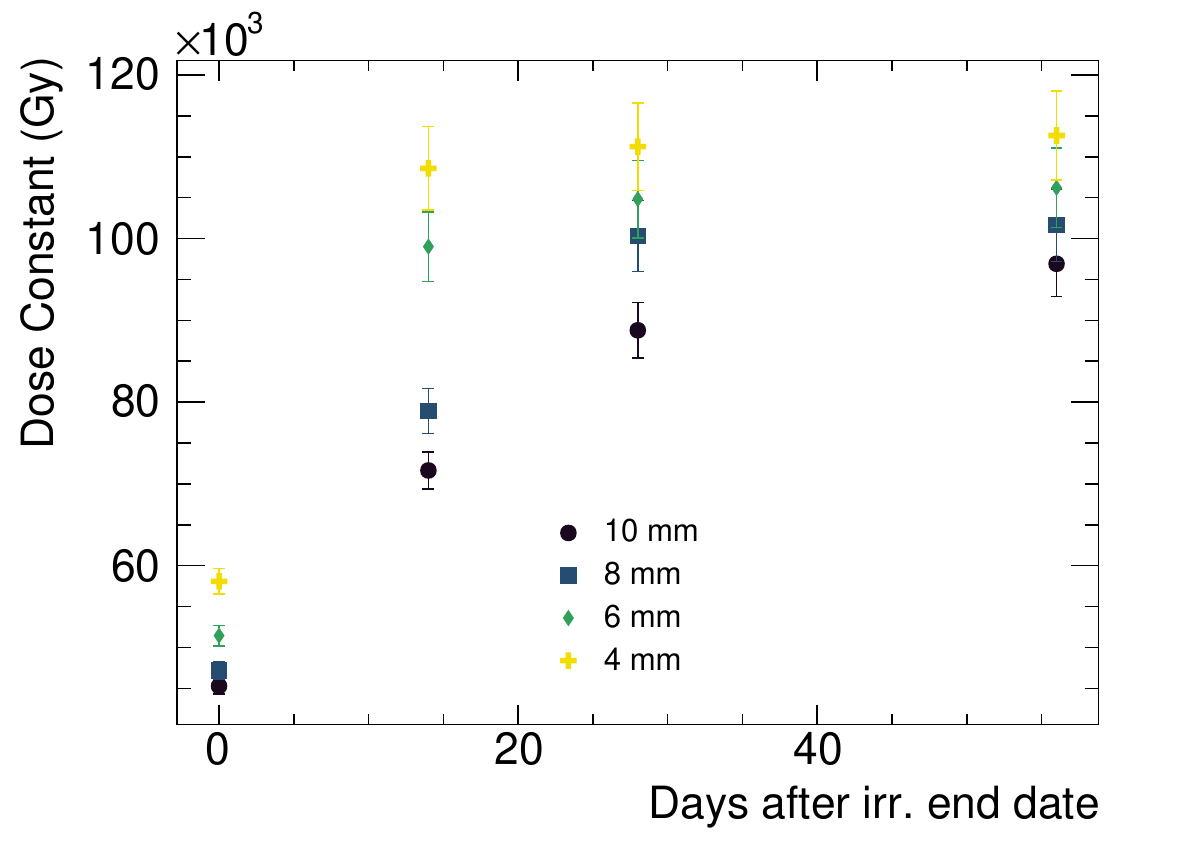}
\includegraphics[width=0.9\textwidth]{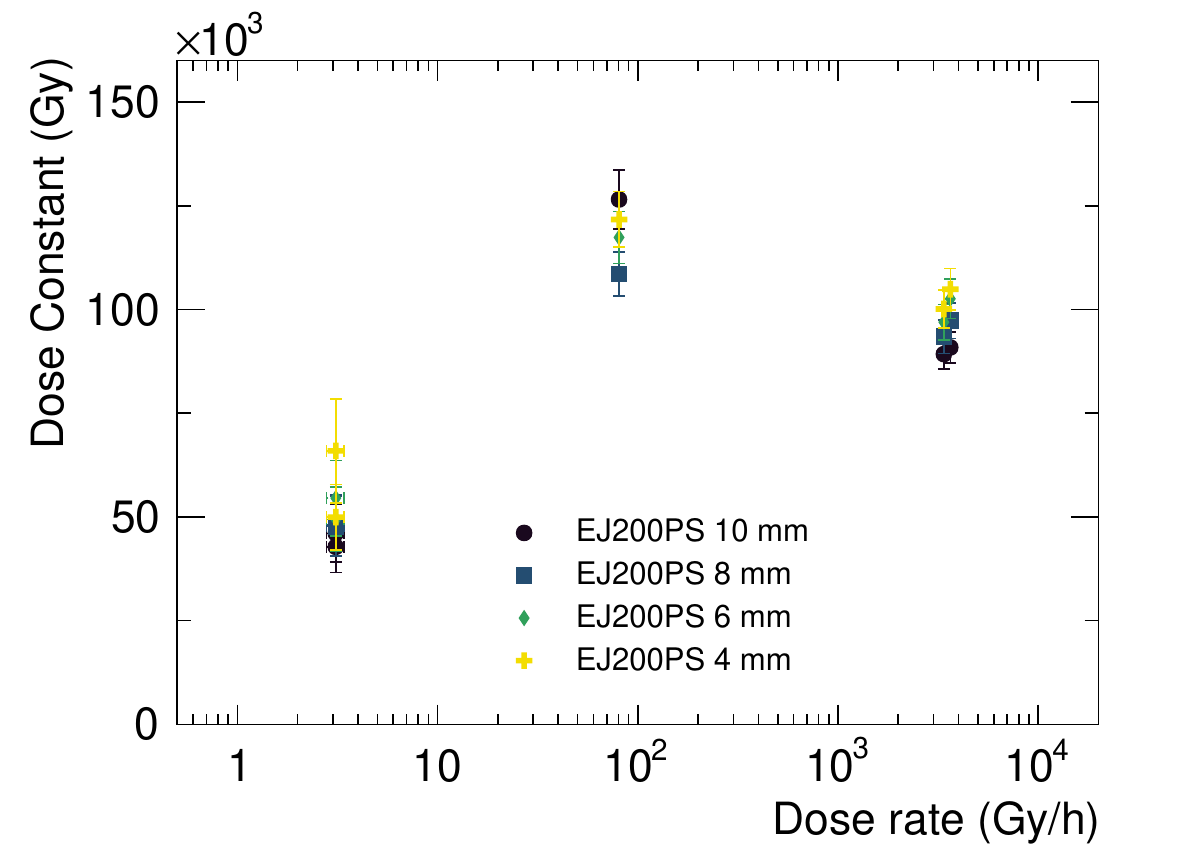}
\caption{For rods with EJ-200 fluors made of PS: [top] value of $D$ versus day since end of irradiation for a $\doserate$ of 3640\unit{Gy/hr}, for rod thicknesses of 0.4\unit{cm}, 0.6\unit{cm}, 0.8\unit{cm}, and 1.0\unit{cm};
[bottom] value of $D$ after annealing versus $\doserate$ for rod thicknesses of 0.4\unit{cm}, 0.6\unit{cm}, 0.8\unit{cm}, and 1.0\unit{cm}. 
}
\label{fig:thick}
\end{figure}

Figure~\ref{fig:thick} shows the results for PS rods with EJ-200 fluors. The results for PVT rods are similar.  The top figure shows $D$ versus date since the end of irradiation.  At 14 days since irradiation, the value of $D$ for the 1.0\unit{cm} rod is 1.5\,$\times$ smaller (more damage) than that of the 0.4\unit{cm} rod, while at 56 days the ratio is 1.1.  The ratio of the thicknesses is 2.5.  Just after irradiation, when the density of color centers is higher, $D$ depends more strongly on the thickness.  After annealing, the dependence is small.  This indicates that for small pieces like this, where the path length in the scintillator is small, after annealing the dominant source of light loss is the change in the initial light production.  The bottom figure shows the value of $D$ after annealing as a function of $\doserate$.  Loss of initial light dominates at all measured $\doserate$s.

\begin{figure}[hbtp]
\centering
\includegraphics[width=0.9\textwidth]{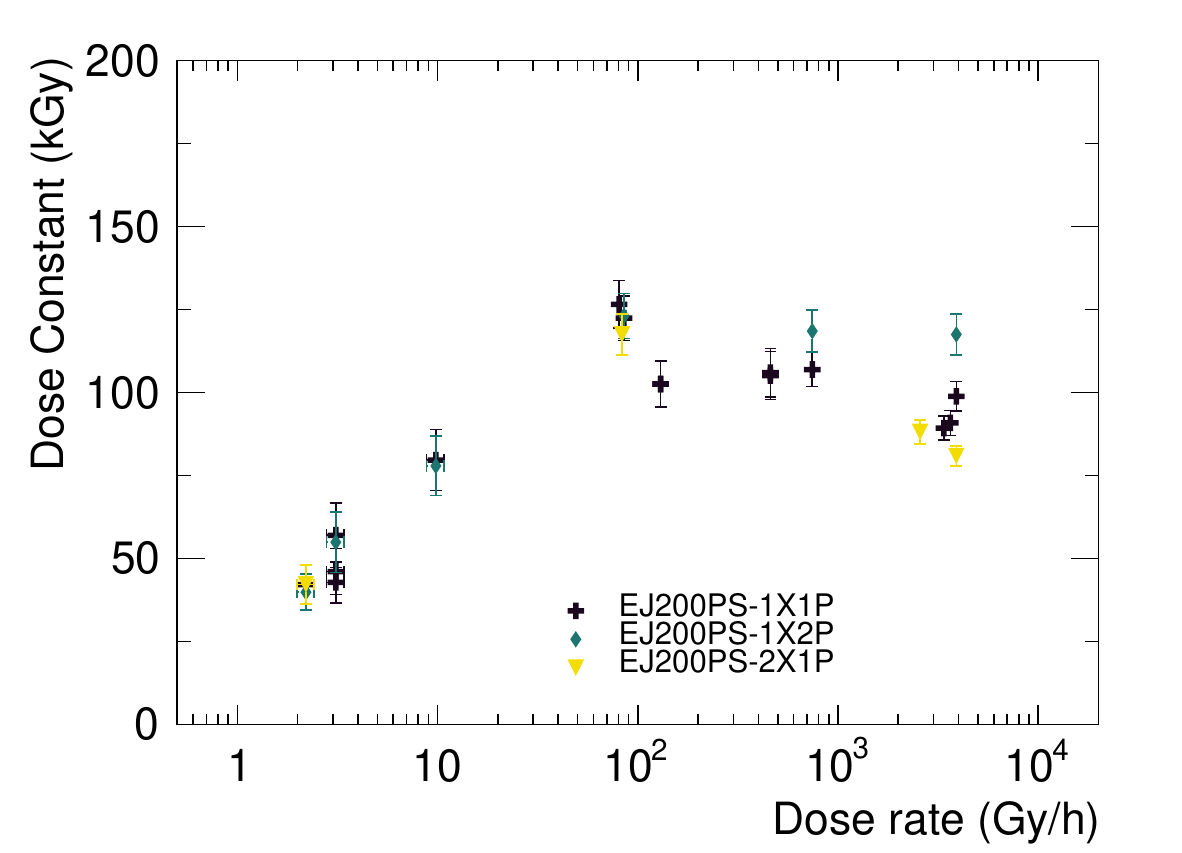}
\includegraphics[width=0.9\textwidth]{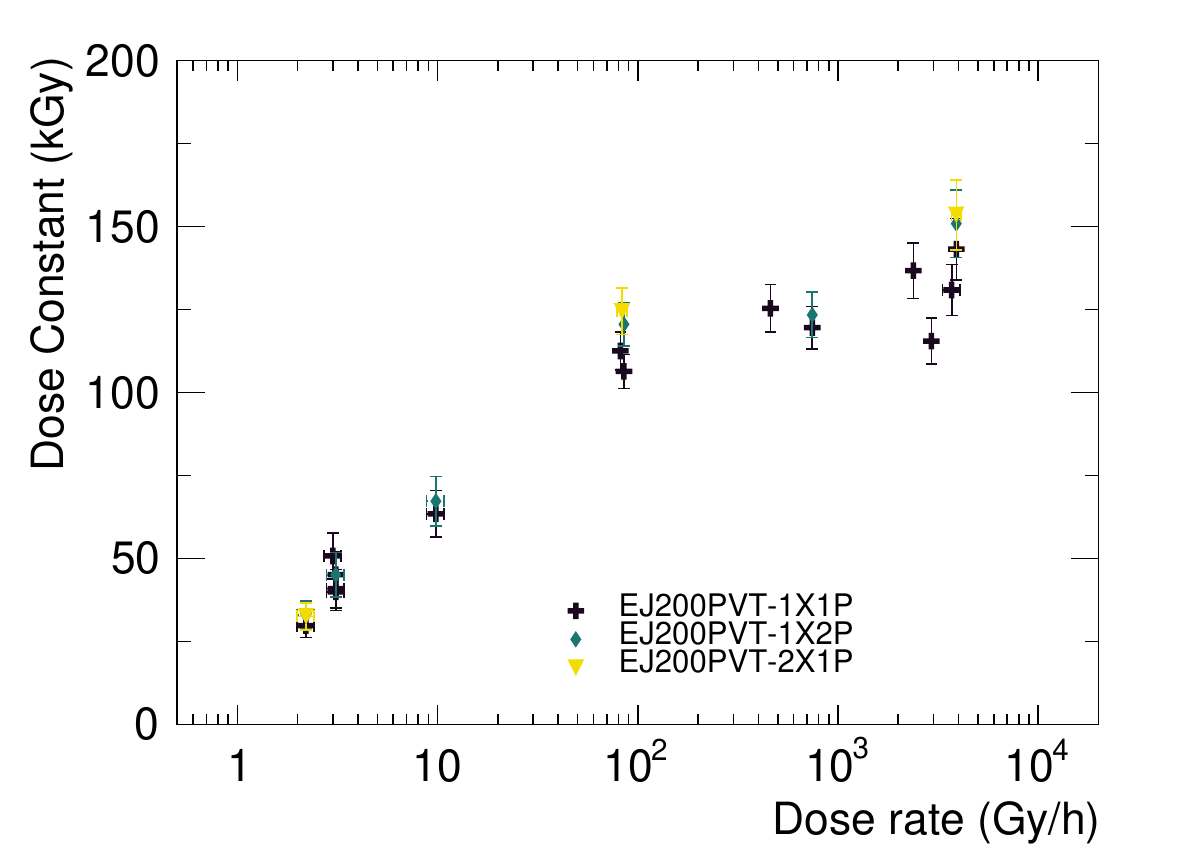}
\caption{[top] The value of $D$ versus $\doserate$ for PS scintillator rods with the EJ-200 fluors at the manufacturer's nominal concentration for antioxidants for the nominal concentration of fluors (1X1P), for double the concentration of the primary fluor (1X2P), and for double the concentration of the secondary fluor (2X1P).
[bottom] The value of $D$ versus $\doserate$ for PVT scintillator rods with the EJ-200 fluors at the manufacturer's nominal concentration for antioxidants for the nominal concentration of fluors (1X1P), for double the concentration of the primary fluor (1X2P), and for double the concentration of the secondary fluor (2X1P).
}
\label{fig:fluors}
\end{figure}

Oxidized polymers tend to absorb light with wavelengths corresponding to emission by the primary fluor.  Because of this, we investigated the effect of varying the dopant and anti-oxidant concentrations. Figures~\ref{fig:fluors} [top] and [bottom] show $D$ versus $\doserate$ for scintillator rods with the EJ-200 fluors, PS and PVT matrix, respectively, and the manufacturer's nominal antioxidant concentration, for nominal fluor concentrations (1X1P), with twice the nominal primary concentration (1X2P) and with twice the nominal secondary concentration (2X1P). The results reveal increased radiation tolerance at high $\doserate$s for PS-based scintillators with twice the nominal primary fluor concentration (PS-1X2P). The rest of the results are independent of the fluor concentration within uncertainties.

\begin{figure}[hbtp]
\centering
\includegraphics[width=0.9\textwidth]{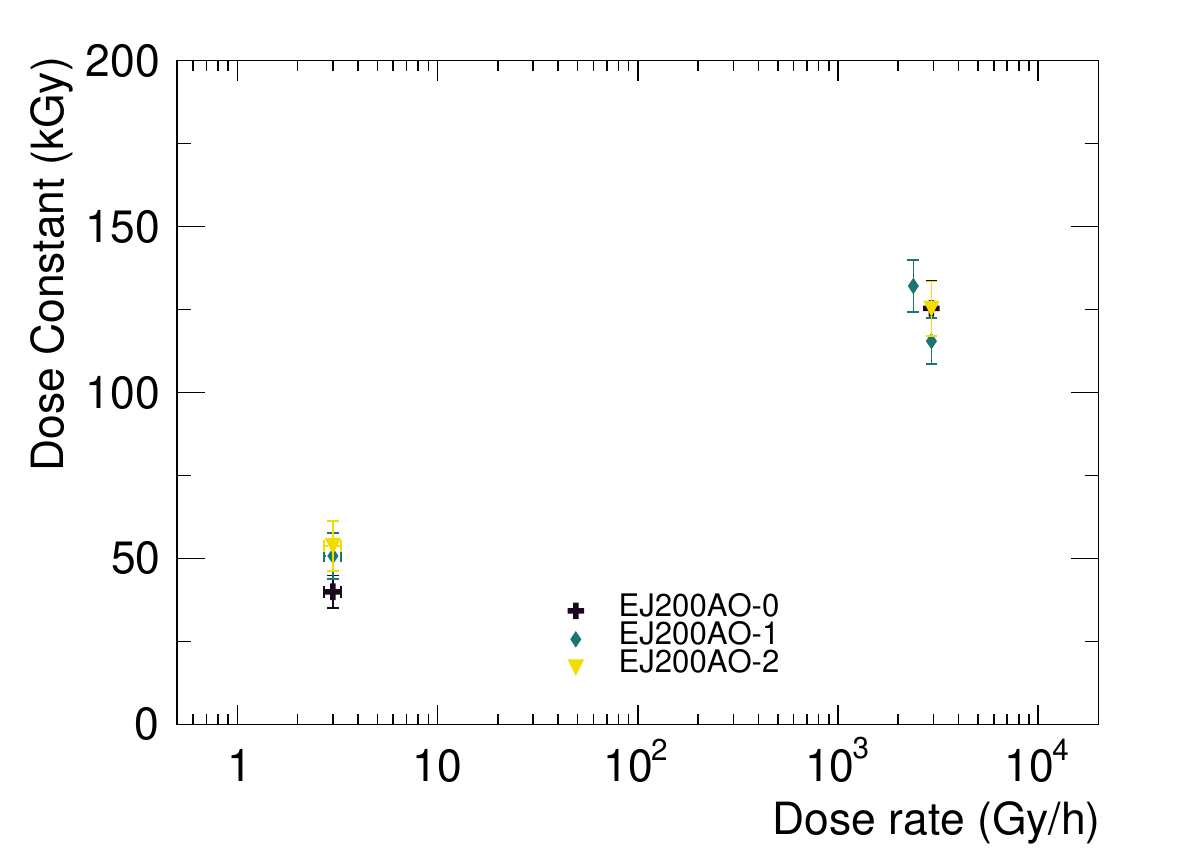}
\caption{The value of $D$ versus $\doserate$ for scintillator rods with the EJ-200 fluors in PVT at the manufacturer's nominal concentration for fluors without antioxidant additives (AO-0), with the nominal concentration (AO-1), and with twice the nominal (AO-2).
}
\label{fig:anti}
\end{figure}

Figure~\ref{fig:anti} shows the results for the EJ-200 fluors at the nominal concentration in PVT without anti-oxidants (AO-0), with the nominal anti-oxidant concentration (AO-1), and with twice the nominal concentration (AO-2).  The results are independent of anti-oxidant concentration within uncertainties. This is expected, since antioxidants are in the amorphous, and not the crystalline regions, of the matrix. PVT is in a glassy state at room temperature.  It is almost impossible for the antioxidant molecules to move around and react with the C-centered radicals or the peroxyl radicals.

\section{Conclusions}

Results on the effects of ionizing radiation on the signal produced by plastic scintillating rods manufactured by Eljen Technology company were presented. Assuming an exponential decrease in the light output with dose, the change in light output was quantified using the exponential dose constant $D$.

The $D$ values are similar for primary and secondary doping concentrations of 1 and 2 times, and for antioxidant concentrations of 0, 1, and 2 times, the default manufacturer's concentration, and so the default concentrations is optimal. The $D$ value depends approximately linearly on the logarithm of the dose rate for dose rates between 2.2\unit{Gy/hr} and 100\unit{Gy/hr} for all materials. For EJ-200 polyvinyltoluene-based (PVT) scintillator, the dose constant is approximately linear in the logarithm of the dose rate up to 3900 Gy/hr, while for polystyrene-based (PS) scintillator or for both materials with EJ-260 fluors, it remains constant or decreases (depending on doping concentration) above about 100 Gy/hr. 

These results show that the conventional wisdom that PVT is more radiation-resistant than PS is only true at high dose rates, for the short paths to the photodetector tested here.  Shifting the wavelength from blue to green again only aids in radiation resistance for dose rates higher than typically found at collider detectors for short path lengths.

The results from rods of varying thickness and from the different fluors suggest damage to the initial light output is a larger effect than color center formation for scintillator thickness $\leq1$~cm. For the blue scintillator (EJ-200), the transmission measurements indicate damage to the fluors. 

\section{Acknowledgments}
The authors would like to thank Chuck Hurlbut of Eljen Technology company for supplying many of the rods and for advice, and the staffs at Goddard Space Flight Center, the National Institute of Standards and Technology, and at the Sandia National Laboratories Facilities group for assistance with the irradiations. We thank Professor Sally Seidel and her group at the University of New Mexico for extensive help with interfacing with Sandia. This work was supported in part by U.S. Department of Energy Grant DESC0010072.

\section*{References}

\bibliography{main}

\newpage
\appendix
\section{List of scintillator samples}\label{app:A}
This appendix contains lists of the samples used throughout the paper.  For the samples with varying fluor concentrations, the scintillator label in the first column has the form ``$<$scintillator color$><$substrate$>$". The scintillator color can be either ``EJ200" or ``EJ260". The substrate can be PS or PVT, where PS in the name means the matrix is polystyrene, and PVT means it is polyvinyltoluene. The fluor concentrations are shown as ``NXMP", where ``N" refers to the multiplication factor for the secondary fluor and ``M" to the primary fluor. For the samples with varying antioxidant concentrations, the scintillator label in the first column has the form ``$<$scintillator color$><$antioxidant concentrations$>$", where antioxidant concentrations are expressed by numbers 0, 1, 2 which represent 0.5, 1, and 2 times the nominal value. 

\begin{table}[h]
\footnotesize
\centering
\begin{tabular}{ccccc}
\toprule 
Scintillator type & Dose (kGy) & Dose rate $\doserate$ (Gy/hr) & Irr. end date & Irr. facility \\\midrule
\multirow{15}{*}{EJ200PS-1X1P}
    & 13.2 &    2.2 & 2018/10/26 & GIF++ \\ 
    & 12.6 &    3.1 & 2018/11/30 & GSFC REF\footnotemark[8] \\ 
    & 42.0 &    9.8 & 2018/12/06 & GSFC REF \\ 
    & 70.0 &   80.6 & 2017/04/12 & NIST \\ 
    & 70.0 &   85.3 & 2016/11/09 & NIST \\ 
    & 69.0 &  130.0 & 2021/09/24 & NIST\footnotemark[9] \\ 
    & 70.0 &  460.0 & 2021/10/14 & NIST\footnotemark[9] \\ 
    & 70.0 &  744.0 & 2016/12/02 & NIST \\ 
    & 70.0 & 3380.0 & 2017/11/07 & NIST \\ 
    & 70.0 & 3640.0 & 2017/04/13 & NIST \\ 
    & 70.0 & 3900.0 & 2016/10/04 & NIST \\ 
\midrule
\multirow{6}{*}{EJ200PS-1X2P} 
    & 13.2 &    2.2 & 2018/10/26 & GIF++ \\ 
    & 12.6 &    3.1 & 2018/11/30 & GSFC REF \\ 
    & 42.0 &    9.8 & 2018/12/06 & GSFC REF \\ 
    & 70.0 &   85.3 & 2016/11/09 & NIST \\ 
    & 70.0 &  744.0 & 2016/12/02 & NIST \\ 
    & 70.0 & 3900.0 & 2016/10/04 & NIST \\ 
\midrule
\multirow{4}{*}{EJ200PS-2X1P} 
    & 13.2 &    2.2 & 2018/10/26 & GIF++\\ 
    & 70.0 &   83.4 & 2017/01/11 & NIST \\ 
    & 70.0 & 2570.0 & 2020/02/14 & NIST \\ 
    & 70.0 & 3900.0 & 2016/10/05 & NIST \\ 
\bottomrule
\end{tabular}
\caption{List of EJ200PS samples.}
\label{tab:samples1}
\end{table}
\begin{table}[h]
\footnotesize
\centering
\begin{tabular}{ccccc}
\toprule 
Scintillator type & Dose (kGy) & Dose rate $\doserate$ (Gy/hr) & Irr. end date & Irr. facility \\\midrule
\multirow{15}{*}{EJ200PVT-1X1P} 
    & 13.2 &    2.2 & 2018/10/26 & GIF++ \\ 
    & 17.1 &    3.0 & 2019/12/04 & GSFC REF \\ 
    & 12.6 &    3.1 & 2018/11/30 & GSFC REF\footnotemark[8] \\ 
    & 42.0 &    9.8 & 2018/12/06 & GSFC REF \\ 
    & 70.0 &   81.9 & 2017/03/01 & NIST \\ 
    & 70.0 &   85.3 & 2016/11/09 & NIST \\ 
    & 70.0 &  744.0 & 2016/12/02 & NIST \\ 
    & 70.0 & 2380.0 & 2020/09/17 & NIST\footnotemark[9] \\ 
    & 60.0 & 2930.0 & 2019/02/22 & NIST \\ 
    & 70.0 & 3700.0 & 2017/03/02 & NIST \\ 
    & 70.0 & 3900.0 & 2016/10/04 & NIST \\ 
\midrule
\multirow{6}{*}{EJ200PVT-1X2P} 
    & 13.2 &    2.2 & 2018/10/26 & GIF++ \\ 
    & 12.6 &    3.1 & 2018/11/30 & GSFC REF \\ 
    & 42.0 &    9.8 & 2018/12/06 & GSFC REF \\ 
    & 70.0 &   85.3 & 2016/11/09 & NIST \\ 
    & 70.0 &  744.0 & 2016/12/02 & NIST \\ 
    & 70.0 & 3900.0 & 2018/10/04 & NIST \\ 
\midrule
\multirow{3}{*}{EJ200PVT-2X1P} 
    & 13.2 &    2.2 & 2018/10/26 & GIF++ \\ 
    & 70.0 &   83.4 & 2017/01/11 & NIST \\ 
    & 70.0 & 3900.0 & 2016/10/05 & NIST \\ 
\bottomrule
\end{tabular}
\caption{List of EJ200PVT samples.}
\label{tab:samples2}
\end{table}
\begin{table}[h]
\footnotesize
\centering
\begin{tabular}{ccccc}
\toprule 
Scintillator type & Dose (kGy) & Dose rate $\doserate$ (Gy/hr) & Irr. end date & Irr. facility \\\midrule
\multirow{6}{*}{EJ260PS-1X1P} 
    & 13.2 &    2.2 & 2018/10/26 & GIF++ \\ 
    & 12.6 &    3.1 & 2018/11/30 & GSFC REF \\ 
    & 42.0 &    9.8 & 2018/12/06 & GSFC REF \\ 
    & 70.0 &   83.4 & 2017/01/11 & NIST \\ 
    & 70.0 &  460.0 & 2020/12/20 & NIST \\ 
    & 70.0 & 3900.0 & 2016/10/05 & NIST \\ 
\midrule
\multirow{6}{*}{EJ260PVT-1X1P} 
    & 13.2 &    2.2 & 2018/10/26 & GIF++ \\ 
    & 12.6 &    3.1 & 2018/11/30 & GSFC REF \\ 
    & 42.0 &    9.8 & 2018/12/06 & GSFC REF \\ 
    & 70.0 &   83.4 & 2017/01/11 & NIST \\ 
    & 70.0 &  460.0 & 2020/12/20 & NIST \\ 
    & 70.0 & 3900.0 & 2016/10/05 & NIST \\ 
\bottomrule
\end{tabular}
\caption{List of EJ260PS and EJ260PVT samples.}
\label{tab:samples3}
\end{table}
\begin{table}[h]
\footnotesize
\centering
\begin{tabular}{ccccc}
\toprule 
Scintillator type & Dose (kGy) & Dose rate $\doserate$ (Gy/hr) & Irr. end date & Irr. facility \\\midrule
\multirow{2}{*}{EJ200AO-0} 
    & 17.1 &  3.0 & 2019/12/04 & GSFC REF \\ 
    & 60.0 & 2930 & 2019/02/22 & NIST \\ 
\midrule
\multirow{3}{*}{EJ200AO-1} 
    & 17.1 &  3.0 & 2019/12/04 & GSFC REF \\ 
    & 70.0 & 2380 & 2020/09/17 & NIST \\ 
    & 60.0 & 2930 & 2019/02/22 & NIST \\ 
\midrule
\multirow{2}{*}{EJ200AO-2} 
    & 17.1 &  3.0 & 2019/12/04 & GSFC REF \\ 
    & 60.0 & 2930 & 2019/02/22 & NIST \\ 
\bottomrule
\end{tabular}
\caption{List of varying antioxidant samples.}
\label{tab:samples4}
\end{table}

\footnotetext[8]{There are 3 samples for this irradiation.}
\footnotetext[9]{There are 2 samples for this irradiation.}

\end{document}